\documentstyle[12pt]{article}
\newcommand{\be}{\begin{equation}}
\newcommand{\en}{\end{equation}}
\newcommand{\bea}{\begin{eqnarray}}
\newcommand{\ena}{\end{eqnarray}}

\newcommand{\hbo}{\hbox to 1 true cm {\hfill } }
\newcommand{\tr}{\hbox{tr}}
\newcommand{\Tr}{\hbox{Tr}}

\def\dslash{\partial\kern-.5em\slash}
\def\kslash{k\kern-.5em\slash}
\def\pslash{p\kern-.5em\slash}
\def\bs{\indent\indent}

\begin{document}
\vglue 1truecm

\vbox{ T95/070
\hfill May 31, 1995
}
\vbox{
hep-ph/9506265
}

\vfil
\centerline{\large\bf Quark Confinement in a Constituent Quark Model}

\bigskip
\centerline{ Kurt Langfeld$^{*}$ and Mannque Rho }
\vspace{1 true cm}
\centerline{\it Service de Physique Th\'eorique, C.E.A. Saclay, }
\centerline{\it F--91191 Gif--sur--Yvette Cedex, France. }
\centerline{and}
\centerline{\it Institute for Nuclear Theory}
\centerline{\it University of Washington}
\centerline{\it Seattle, WA 98195, U.S.A.}
%
\vskip 1.5cm

\begin{abstract}
\noindent
On the level of an effective quark theory, we define confinement
by the absence of quark anti-quark thresholds in correlation functions.
We then propose a confining Nambu-Jona-Lasinio-type model. The confinement
is implemented in analogy to Anderson localization in condensed matter
systems. We study the model's phase structure as well as its behavior
under extreme conditions, i.e.\ high temperature and/or high density.

\end{abstract}

\vfil
\hrule width 5truecm
\vskip .2truecm
\begin{quote}
$^*$ Supported in part by DFG under contracts La--$932/1-1/2$.
\end{quote}
\eject
\section{ Introduction }
\label{sec:1}
\bs

The description of hadron physics starting from quantum chromodynamics
\break
(QCD) -- the theory of strong interactions -- is one of the
most challenging problems in medium energy physics of today.
The difficulty in the description of the low energy sector of QCD
lies in the large effective coupling constant, which precludes
a perturbative treatment.

Various non-perturbative methods
shed some light on its ground state properties: the operator product
expansion~\cite{wi69} improves the perturbative expansion by the
inclusion of effects from vacuum condensates in Green's functions,
QCD sum rules~\cite{shi79} relate these condensates to hadron masses,
variational methods~\cite{ole81} attempt to evolve a qualitative
picture of the gluonic vacuum, the semi-classical evolution of the
QCD functional integral, leading to instanton physics~\cite{ca78},
and the reformulation of QCD in terms of the field
strength~\cite{ha77,sch90,re91,la91} provides some insight into the vacuum
properties
of the quark sector. These analytical approaches can be contrasted to
numerical investigations of lattice QCD~\cite{wi74,eng87}. The latter
approach is not bounded by any approximation, but is restricted
by the capacities of computers.

One of the most striking features
of low energy QCD is the absence of quarks in the asymptotic states.
This quark confinement is explained in lattice QCD by
a linear rising potential between two static quarks. This behavior
is also confirmed by considerations within the $1/N_c$-expansion,
$N_c$ being the number of colors~\cite{mi83}. A natural explanation
of the linear confining potential is provided by the dual superconductor
picture~\cite{ba91}. As pointed out by 't Hooft, in a certain gauge,
the non-abelian Yang-Mills theories possess monopole
configurations~\cite{tHooft}. If these monopoles condense, a dual
Meissner effect occurs, expelling electric field strength out of the
vacuum. This implies that the electric field between two static
quarks is squeezed into a flux tube, which subsequently gives rise
to the linear confining potential. A recent SUSY model
of Seiberg and Witten
suggests that monopole condensation could be the mechanism of confinement
in certain Yang-Mills theories~\cite{sei94,sei94b}. There have
been suggestions that the same phenomenon occurs in the Yang-Mills sector
of QCD proper \cite{qcd}. In this development, it is
found that the confinement of the quarks is intimately related
to the spontaneous breakdown of chiral symmetry~\cite{sei94b}.

Despite the recent progress in understanding the ground state properties
of QCD, the description of hadron properties is still feasible only with
effective models. These models include aspects of QCD by incorporating
its symmetries. The most important symmetry constraining the variety of
hadron models is chiral symmetry. For QCD ($N_c=3$) and for
zero current masses, the quark sector is invariant under global
$SU(N_f)_L \times SU(N_f)_R$ transformations of the left- and right-handed
quarks, where $N_f$ is the number of quark flavors. The vaccum breaks
this symmetry down to the diagonal $SU(N_f)$.\footnote{For $SU(2)$
Yang-Mills theory, the chiral symmetry group is much larger,
i.e.\ $SU(2N_f)$, since the $SU(2)$ gauge group is pseudoreal~\cite{sm95}.
More on this later.}

Most effective theories of mesons and nucleons
contain many parameters restricting the predictive power of these
models. It is therefore desirable to ``derive" the effective hadron theory
from an underlying effective quark model in order to obtain constraints
on the parameter range. One of these quark models is the model of
Nambu and Jona-Lasinio (NJL)~\cite{nambu}. The NJL model is one of the most
economical models that possess the essence of dynamical symmetry breaking
that is the hall-mark of modern field theories and has enjoyed an impressive
phenomenological success in hadron physics~\cite{appli}.
Now the NJL-model reflects the principal low-energy
QCD symmetry properties of the quark sector, but does not include
the confinement of quarks. This implies, in particular, that the
mesons of this model -- quark-anti-quark bound states --
can decay into free quark-anti-quark pairs, if such a process
is allowed by kinematics. This manifestly unphysical threshold puts severe
constraints on the applicability of the NJL-model~\cite{ja92}.

In this paper, we propose an NJL-type model that possesses the confinement
property in the sense that quark-anti-quark thresholds are absent
in (mesonic) Green's functions.
Our reasoning, albeit exploratory,
 will be guided by a close analogy to a phenomenon
in condensed-matter physics known as Anderson localization
of electrons~\cite{anderson}. In Anderson localization, it is observed
that freely moving
electrons get localized when the strength of a random distributed
potential stemming from the impurities of the conducting solid exceed a
certain strength. We shall argue in analogy that
the quarks feel randomly distributed background fields generated by
the gluon sector.

The paper is organized as follows. In the next section,
a motivation and a detailed
description of the model is presented. The gap equation describing
quark ground-state properties is derived. In section 3,
we first describe the particular ansatz which produces a remarkable
confinement property. The absence of quark-anti-quark thresholds is
explicitly demonstrated for the scalar correlation function. We then
show that this particular ansatz is indeed a solution of the gap equation.
The phase structure of the model is discussed in some detail.
In section 4, we describe the implication of the model on
chiral properties. There is no surprise here as the model is
designed to reproduce the chiral structure of QCD (section 2).
We reproduce the corresponding low-energy theorems by first
solving the Bethe-Salpeter equation for the pion field, normalize
the amplitude by the electromagnetic form factor and extract the pion
decay constant from its definition. We then establish that the
Gell-Mann-Oakes-Renner (GMOR) relation is valid in our model. Finally, we
compute the pseudoscalar correlation function. The sections 5 and 6 are
devoted to the temperature and density dependence of the vacuum properties
of the model. We will show that a deconfinement phase transition occurs at
some large temperature and/or density. We will find that the
deconfinement phase transition is accompanied by the restoration of the
spontaneously broken chiral symmetry. The model predicts the two
transitions at the same critical point. The final section contains
some concluding remarks.

\section{The Model }
\label{sec:2}

\subsection{ Motivation }

\bs

It is known for a long time that the lattice formulation of
Yang-Mills theory provides information on
the confinement of quarks~\cite{eng87}.
Subsequently quark liberation due to temperature was studied with
the lattice version of QCD. It turned out that quantities which
are dominated by the properties of the gluonic sector, e.g.\ the
gluon condensate, vary smoothly throughout the deconfinement phase
transition \cite{lattice,kochbrown}.
This suggests that quark liberation may be described solely by a dynamical
effect of the quark sector. If so,
the description of this phase transition should be feasible by an
effective quark model. We hope to gain qualitative insight into the
nature of quark liberation by including temperature effects in the
quark loops, while the temperature dependence of the quark
interaction, mediated by the gluonic sector, is kept temperature-independent.

The properties of the light pseudoscalar mesons, in particular those of
the scalar meson and the pion, are {\it protected } by chiral symmetry
of QCD and might therefore serve as a convenient test ground to explore
other features of low energy QCD. We have learned in the past years
that at low energies, the pion physics is
phenomenologically well described in terms of an effective quark model,
in which the quarks interact via a local current-current interaction
of the NJL type~\cite{nambu,appli}. However, the open question is:
What is the
signature of confinement at the level of an effective quark theory?

In order to answer this question, we can be guided by a phenomenon in
solid state physics: the localization of electrons in random
potentials~\cite{anderson}, known as Anderson localization. In his
pioneering work, Anderson showed that the key idea of localization
can be traced back to the Hamiltonian
\be
H \; = \; \sum _{ \langle ik \rangle } (-V) \, a_{i}^{\dagger }
a_{k} \; + \; \sum _i \epsilon _i \, a_{i}^{\dagger } a_i \; ,
\label{eq:2.1.1}
\en
where $a_{i}^\dagger $ is the creation operator of a spin at site
$i$ of a lattice. The sum in the first term of (\ref{eq:2.1.1}) extends
only over the nearest neighbors, and  the constant $V$ measures the strength
of the nearest-neighbor hopping. This term is responsible for the spin
diffusion on the lattice and is the analogue of a kinetic term in
continuum quantum field theory. The energy $\epsilon _i $ corresponds
to a potential at site $i$. In the Anderson model, the $\epsilon _i$'s
are random variables distributed over a range $-W/2 < \epsilon <
W/2 $. A continuum version of this model is described by the Hamiltonian
\be
H \; = \; p^2 \; + \; w \, \sum _{\alpha } \delta ( r - R_\alpha ) \, ,
\label{eq:2.1.2}
\en
which describes electrons elastically scattered off impurities which
are randomly distributed at points $R_\alpha $.
It turns out that the precise form of the distribution of the
potential $\epsilon _i$ is not crucial. It has been found that if the
random potential exceeds a certain strength, i.e.
\be
W \; > \; W_c \, = \, 4 K V \; ,
\label{eq:2.1.3}
\en
where the connective constant $K$ is characteristic of the lattice type, then
the spins are localized at their sites, whereas for $W<W_c$ the spins
are {\it liberated }, and the spin diffusion takes place.

The localization of spins is intimately related to the presence of
random potentials. Is there a random interaction of quarks which
stems from the gluonic sector and which survives the low-energy limit?
The answer to this question is not known. There is, however,
a first hint from the field-strength formulation of QCD~\cite{ha77,sch90}.
When QCD ($N_c=3$) is formulated in terms of the field strength, the resulting
quark interaction takes the form
\be
Z[j] \; = \; \int {\cal  D } T^{a}_{\mu \nu } \; e^{-S[T]} \;
\exp \left\{ \int d^4x \; \left[ j^{a}_{\mu } V^{a}_{\mu }
\, -i \frac{g}{2} j^{a}_{\mu } (\hat{T}^{-1} )^{ab}_{\mu \nu }
j^{b}_{\nu } \right] \right\} \; ,
\label{eq:2.1.4}
\en
where $j^{a}_{\mu } = \bar{q} t^a \gamma _\mu q$ is the color octet
current of the quarks, $\hat{T} ^{ab}_{\mu \nu } = f^{abc}
T^{c}_{\mu \nu }$ is defined with the help of the $SU(3)$ structure functions
$f^{abc}$, and $V^{a}_\mu = (\hat{T}^{-1} )^{ab}_{\mu \nu }
\partial _\rho T^{b}_{\rho \nu } $ is the gauge potential induced
by the conjugate field strength $T^{a}_{\mu \nu }$. The action of
the field strength $S[T]$ need not to be specified here and can
be found in~\cite{ha77,sch90}. Within the strong coupling limit, one
observes that the gluonic vacuum decays into domains of constant
field strength $i T^{a}_{\mu \nu }$~\cite{sch90} giving rise to
a Nambu-Jona-Lasinio type quark interaction in~(\ref{eq:2.1.4}).
Due to gauge covariance, all orientations of the constant background
field contribute to the quark interaction. Subsequently, the
spontaneous breakdown of chiral symmetry was observed in the
strong coupling limit~\cite{re91,la91}. The specific form of the
quark interaction due to the gluon background fields gives rise
to a splitting of the strange- and up-quark condensates as is
predicted by QCD sum rules~\cite{la91}.

We conclude from these observations that the low energy-effective
quark interaction of the NJL type is in fact mediated
by the gluon background field. Our key assumption is that
in strongly fluctuating gluonic background fields (with an average scale
given by the gluon condensate), we will be endowed
with a random quark interaction which induces quark confinement
by a mechanism similar to
Anderson localization in solid state physics. In this paper, we
shall construct a simple toy model which has the expected qualitative
feature mentioned above of low-energy QCD.

\subsection{ Description of the model }
\label{sec:2.2}

\bs
In order to investigate the implications
of strong random colored interactions of quarks, we study
a model in which the quark fields are doublets of a global $SU(2)$
{\it color } symmetry.
We write the generating functional for mesonic Green's functions
in Euclidean space as
\bea
Z[\phi ] &=& \left\langle \int {\cal D} q \; {\cal D } \bar{q} \;
\exp \left\{ - \int d^{4}x \; [
{\cal L } \, - \, \bar{q}(x) \phi (x) q(x)] \; \right\}
\right\rangle _{ O } \; ,
\label{eq:1} \\
{\cal L }  &=& \bar{q}(x) ( i \dslash + im ) q(x) \; + \; \frac{G_0}{2}
[ \bar{q} q(x) \, \bar{q} q(x) \, - \, \bar{q} \gamma _5 q(x) \,
\bar{q} \gamma _5 q(x) ]
\label{eq:2} \\
&+& \frac{1}{2} [ \bar{q} \tau ^{\alpha } q(x)  \, G^{\alpha \beta }
\bar{q} \tau ^{\beta } q(x) \, - \,
\bar{q} \gamma _5 \tau ^{\alpha } q(x)  \, G^{\alpha \beta }
\bar{q} \gamma _5 \tau ^{\beta } q(x)  ] \; ,
\nonumber
\ena
where $m$ is the current quark mass. We assume that
the quark interaction is given by a color-singlet four-fermion
interaction of strength $G_0$ and a color-triplet interaction
mediated by a positive definite matrix $G^{\alpha \beta }$,
which represents gluonic background fields. An average
of all orientations $O$ of the background field $G^{\alpha \beta }$,
transforming as $G'=O^{T} G O$ with $O$ being a $3 \times 3$
orthogonal matrix, is understood in (\ref{eq:1}) to
restore global $SU(2)$ color symmetry.  Our basic assumption as
motivated above is that this
averaging amounts to assuring confinement.

Our model is defined in Euclidean space for two reasons.
First, the motivation of the model is provided by the Euclidean
formulation of QCD. It seems reasonable to assume that
classical Euclidean configurations
such as instantons or monopoles might provide the random background.
Second, the averaging procedure in (\ref{eq:1}) is better defined in
Euclidean space, since a superposition of weight functions is again
a weight function (with a correct normalization), whereas
the superposition of phases (as the integrand of the Minkowskian
functional integral is) does not give a phase. The theory in
Minkowski space {\it is defined } by the standard Wick rotation.
We will have more discussions on this later.

Instead of the current-current interaction (\ref{eq:2.1.4}),
we shall use its pseudoscalar-scalar
part that results from a Fierz transformation. The reason is as follows.
As mentioned, it is known that QCD in $SU(2)$ (being pseudoreal) undergoes
a symmetry breakdown which is quite different from what one expects
in three-color QCD. To simulate what happens in QCD with three colors,
we choose interactions so that we would have the correct
symmetry-breaking pattern. Specifically
whereas the interaction of two color triplet currents $j_\mu ^a$
which presumably follows from QCD exhibits the
full $SU(2N_f)$ chiral symmetry group, the reduced interaction in
(\ref{eq:1}) is, however, only invariant under $SU(N_f) \times SU(N_f)$
transformation, as QCD is. Although we are dealing with an $SU(2)$ color
group, we assume that the basic idea of the confinement mechanism,
developed below, does not depend qualitatively on the color group
under investigations\footnote{
We do not expect the answer to the question as to whether
the quarks are confined or not to depend on the color group used,
whereas the actual value of e.g.\ the pion decay constant might depend
on whether we have the $SU(3)$ or the $SU(2)$ gauge group.}.
In order to keep contact with QCD as closely as
possible, we therefore choose our low-energy effective quark theory to
exhibit the chiral patterns of QCD in four dimensions.

In order to make  contact with more familiar formulations of effective
quark models, we first study the limit where the color-triplet
interaction $G^{\alpha \beta }$ is weak. In this limit we can
perform the average over the background orientation $O$ using
a cumulant expansion. The colored part of the quark interaction becomes
\be
\int d \lambda \; f(\lambda ) \exp \left\{
- \frac{\lambda }{3} \int d^4x \;
[ \bar{q} \tau ^{\alpha } q(x)  \, \bar{q} \tau ^{\alpha } q(x)
\, - \, \bar{q} \gamma _5 \tau ^{\alpha } q(x)  \,
\bar{q} \gamma _5 \tau ^{\alpha } q(x)  ] \; + \; O(\lambda ^2)
\right\} \; ,
\label{eq:3}
\en
where $\lambda $ is an eigenvalue of the matrix $G^{\alpha \beta }$,
and $f(\lambda )$ is the corresponding eigenvalue density.
At lowest order of the cumulant expansion, we obtain the
familiar Nambu-Jona-Lasinio model~\cite{nambu} with a global color
symmetry. Terms multiplied by $\lambda ^2$ are eight-quark interactions.
The average over the background field $G^{\alpha \beta }$
in (\ref{eq:1}) obviously incorporates the interaction of more than
four quarks. If these interactions are not small or equivalently
if the background fields in (\ref{eq:1}) are not weak,
one then has to abandon the cumulant expansion, and instead study the
quark theory of (\ref{eq:1}) with fixed background field in a certain
approximation, which we will specify below, and average over the
background fields afterwards. If the approximation used is not bound
to weak couplings, this approach can be applied even if the cumulant
approximation fails.

Specifically, the approximation under investigation is
to introduce meson fields on top of the  scalar condensate of the vacuum
and to treat their interactions perturbatively. This approximation
does not resort to small couplings and is expected to give good
results if the number of mesonic degrees of freedom is large.
This approximation is the usual one applied to study the physics of
light hadrons in the context of the Nambu-Jona-Lasinio
model. We will not further question the validity of the
approximation, but investigate the ground state and the properties of
the light mesons within this scheme. The quark interaction in (\ref{eq:1})
is linearized by means of color-triplet ($\sigma ^{\alpha }$, $\pi ^{\alpha }$)
and color-singlet mesons ($\sigma $, $\pi $). Integrating out the
quark fields in the Hubbard-Stratonovich formalism,
the resulting effective meson theory is
\bea
Z[\phi ] &=& \left\langle \int {\cal D} \sigma ^{\alpha } \;
{\cal D} \pi ^{\alpha } \; {\cal D} \sigma \; {\cal D} \pi \;
\exp \left\{ - \int d^{4}x \; {\cal L }_M  \; \right\}
\right\rangle _{ O } \; ,
\label{eq:4} \\
{\cal L }_M  &=& - \ln \left( i \dslash + iM_0 + i\sigma + i
\tilde{M}^{\alpha } \tau ^{\alpha } + i \sigma ^{\alpha } \tau ^{\alpha } +
\pi \gamma _5 + \pi ^{\alpha } \tau ^{\alpha } \gamma _5 \right)
\label{eq:5} \\
&+& \frac{1}{2} \left( \sigma ^{\alpha } + \tilde{M}^{\alpha } \right)
\, \left( G^{-1} \right) ^{\alpha \beta }
\left( \sigma ^{\beta } + \tilde{M}^{\beta } \right) \, + \,
\frac{1}{2} \pi ^{\alpha } \left( G^{-1} \right) ^{\alpha \beta }
\pi ^{\beta }
\nonumber \\
&+& \frac{1}{2 G_0 } \left( (\sigma + M_0 - m - i \phi _0)^2 +
( \pi  + \phi _5) ^2 \right) \; ,
\nonumber
\ena
where $\tilde{M}^\alpha $ and $M_0$ are, respectively,
the color-triplet and color-singlet constituent quark masses,
and we have
decomposed the external source $\phi $ into a scalar and pseudoscalar
parts, i.e. $\phi = \phi _0 + \phi _5 \gamma _5 $.
The approximation mentioned above consists of truncating the expansion
of (\ref{eq:4}) in terms of the meson fields. The zeroth-order
approximation provides access to ground state properties, in particular,
the quark condensate. From
\be
\frac{ \delta \, \ln Z[0] }{ \delta \sigma ^{\alpha } } \; = \; 0,  \hbo
\frac{ \delta \, \ln Z[0] }{ \delta \sigma } \; = \; 0
\label{eq:6}
\en
one obtains
\bea
- \frac{1}{V_4} \Tr \left\{ \frac{ i }{ i \dslash + i M_0 + i
\tilde{M}^{\alpha } \tau ^{\alpha } } \right\} &+& \frac{1}{G_0}
\left( M_0 -m \right) \; = \; 0 \; ,
\label{eq:7} \\
- \frac{1}{V_4} \Tr \left\{ \frac{ i }{ i \dslash + i M_0 + i
\tilde{M}^{\alpha } \tau ^{\alpha } } \tau ^{\beta } \right\} &+&
\left( G^{-1}\right )^{\beta \gamma } \tilde{M} ^{\gamma } \; = \; 0 \; ,
\label{eq:8}
\ena
where $V_4$ is the Euclidean space-time volume, and
the trace extends over internal degrees of freedom as well as over
space-time. In the latter case, a regularization is
required, which we will specify when needed.
Different solutions of the equations (\ref{eq:7}-\ref{eq:8})
correspond to different phases of the system.

\section{ The Confining Phase }
\label{sec:3}

\bs

Here we will show that a particular solution of the gap equations
(\ref{eq:7})--(\ref{eq:8}) with an imaginary color-triplet constituent mass
exists, i.e. $\tilde{M} ^\alpha = -i M^{\alpha }$ with $M^{\alpha } $ real.
Before discussing in detail the existence of such a solution,
we first illustrate its remarkable physical consequence.

\newpage
\subsection{ The scalar correlation function }
\label{sec:3.1}

\bs

Consider the color-singlet scalar correlation function
\be
\Delta _s (p) \; = \;
\int d^{4}x \; e^{-ipx} \; \left\langle \bar{q}q(x) \; \bar{q}q (0)
\right\rangle \; ,
\label{eq:9}
\en
within the effective meson theory (\ref{eq:5}). Its
connected part is given by
\be
\Delta _s^c(p) \; = \; \int d^{4}x \; e^{-ipx} \;
\frac{ \delta ^2 \, \ln Z[\phi ] }{ \delta \phi _0 (x)
\delta \phi _0(0) } \vert _{\phi =0 } \; .
\label{eq:10}
\en
To compute this,
one has to consider the fluctuations of the scalar color-singlet
field $\sigma _0$ as well as those of the fields $\sigma ^{\alpha }$,
since, for a fixed interaction $G^{\alpha \beta }$, the colored
mesons couple to $\sigma _0$ by a quark loop (see the first term on the
right-hand side of (\ref{eq:5})). Fluctuations of the pion fields
$\pi ^{\alpha }$, $\pi $ do not contribute to the scalar correlation
function, since the pion fields have the wrong quantum numbers.
Expanding (\ref{eq:5}) up to second order in the meson fields,
the action, in momentum space representation, is
\bea
S^{(2)} &=& \int \frac{ d^4p }{ (2\pi )^4 } \; \left\{
\frac{1}{2} \sigma (p) \Pi _s^0 (p^2) \sigma (-p) +
\frac{1}{2} \sigma ^\alpha (p) \Pi _s^{\alpha \beta } (p^2)
\sigma ^\beta (-p)
\right. \label{eq:e1} \\
&+& \left. i \sigma ^\alpha (p) K^\alpha (p^2) \sigma (-p) -
\frac{ 1 }{ 2 G_0 } \phi _0 (p) \phi _0(-p)
- \frac{i}{ G_0 } \phi _0 (p) \sigma (-p) \, \right\} \; ,
\nonumber
\ena
where
\bea
\Pi _s^0 (p^2) &=& \frac{1}{G_0} \; - \; \int \frac{ d^4k }{ (2\pi )^4 }
\; \tr \left\{ S(k+p) S(k) \right\} \; ,
\label{eq:e2} \\
\Pi _s^{\alpha \beta } (p^2) &=& (G^{-1})^{\alpha \beta }
\; - \; \int \frac{ d^4k }{ (2\pi )^4 }
\; \tr \left\{ \tau ^{\alpha } S(k+p) \tau ^{\beta } S(k) \right\} \; ,
\label{eq:e3} \\
K^{\alpha } (p^2) &=&  -i \; \int \frac{ d^4k }{ (2\pi )^4 }
\; \tr \left\{ \tau ^{\alpha } S(k+p) S(k) \right\} \; .
\label{eq:e4}
\ena
The quark propagator $S(k)$, in momentum space, is
\be
S(k) \; := \; \frac{ 1 }{ \kslash \, + \, i(M_0 + i
M^{\alpha } \tau ^{\alpha } ) } \; .
\label{eq:e5}
\en
Our model (\ref{eq:1}) is defined with an average over all
global $SU(2)$ orientations of the interaction matrix $G^{\alpha \beta }$.
In order to render the averaging procedure easy, we integrate out
the colored mesons in (\ref{eq:e1}), i.e.
\bea
S_{eff}^{(2)} &=& \int \frac{ d^4p }{ (2\pi )^4 } \; \left\{
\frac{1}{2} \sigma (p) \, \left[ \Pi _s^0 \, + \, K^{\alpha }
( \Pi _s ^{-1} ) ^{\alpha \beta } K^{\beta } \right] \,
\sigma (-p)
\right. \label{eq:e6} \\
&-& \left. \frac{1}{ 2 G_0 } \phi _0 (p) \phi _0(-p)
- \frac{i}{ G_0 } \phi _0 (p) \sigma (-p) \, \right\} \; .
\nonumber
\ena
In fact, we will observe that $S_{eff}^{(2)}$ no longer depends
on the orientation of the interaction matrix, so the
averaging procedure is trivial.

In order to study the effective theory of the singlet meson $\sigma $,
we explicitly calculate the polarization functions $\Pi _s^0 $,
$\Pi _s^{\alpha \beta }$ in (\ref{eq:e2}) and (\ref{eq:e3}),
respectively, as well as the mixing $K^{\alpha }$ (\ref{eq:e4}).
For this purpose, it is convenient to introduce the eigenvectors of the
matrix $M^{\alpha } \tau ^{\alpha } $, i.e.
\be
M^{\alpha } \tau ^{\alpha } \, \vert \pm \rangle \; = \;
\pm M \, \vert \pm \rangle \; , \hbo M=\sqrt{ M^{\alpha } M^{\alpha } } \; .
\label{eq:12}
\en
They possess the property
\be
\langle \pm \vert \tau ^\alpha \vert \pm \rangle \; = \;
\pm \frac{ M^\alpha }{M} \; ,
\label{eq:12a}
\en
which will be extensively used later.
The detail of the calculation is left to Appendix \ref{app:a}. It turns out
that the quantities (\ref{eq:e2}) and (\ref{eq:e4}) can be expressed in terms
of two functions $H_0(p^2)$, $H_v(p^2)$, i.e.
\be
\Pi_s^0 (p^2) = \frac{1}{G_0} - H_0(p^2) \; , \hbo
K^{\alpha }(p^2) = \frac{ M^{\alpha } }{ M } \, H_{v}(p^2) \; .
\label{eq:e7}
\en
For later convenience, we explicitly write them down here:
\bea
H_0 (p^2) &=& 4 \int _0^1 d\alpha \; \int \frac{ d^4q }{ (2\pi )^4 } \;
\left\{ \frac{ q^2 -Q }{ \left[ q^2 + Q \right] ^2 } \; + \;
( M \rightarrow -M ) \right\} \; ,
\label{eq:e8} \\
H_v (p^2) &=& -4i \int _0^1 d\alpha \; \int \frac{ d^4q }{ (2\pi )^4 } \;
\left\{ \frac{ q^2 -Q }{ \left[ q^2 + Q \right] ^2 } \; - \;
( M \rightarrow -M ) \right\} \;
\label{eq:e9}
\ena
where
\be
Q \; = \; \alpha (1 - \alpha ) p^2 \; + \; (M_0 + i M)^2 \; .
\label{eq:e10}
\en
We also find (see Appendix \ref{app:a}) that $M^\alpha $ is an
eigenvector of the polarization matrix $\Pi _s^{\alpha \beta }$, i.e.
\be
\Pi _s^{\alpha \beta } \frac{ M^{\beta } }{M} \; = \;
\left( \frac{1}{G_c} - H_0(p^2) \right) \; \frac{ M^{\alpha } }{M} \; ,
\label{eq:e11}
\en
where we have used the property (of which we will have more to say
in the next subsection) of the solution of the Dyson-Schwinger equations
(\ref{eq:7},\ref{eq:8}), namely, that $M^{\alpha }$ is an eigenvector of
the symmetric matrix $G^{\alpha \beta }$ with eigenvalue $G_c$.
It is now straightforward to calculate $S_{eff}^{(2)}$, (\ref{eq:e6}).
Since $K^{\alpha }$ is proportional to $M^{\alpha }$ and using
that the eigenvectors of the symmetric matrix $G^{\alpha \beta }$
are orthogonal, one obtains
\bea
S_{eff}^{(2)} &=& \int \frac{ d^4p }{ (2\pi )^4 } \; \left\{
\frac{1}{2} \sigma (p) \, \left[ \frac{1}{G_0} - H_0 (p^2) \, + \,
\frac{ H_v^2(p^2) }{ \frac{1}{G_c} - H_0(p^2) }
\right] \sigma (-p)
\right. \label{eq:e12} \\
&-& \left. \frac{1}{ 2 G_0 } \phi _0 (p) \phi _0(-p)
- \frac{i}{ G_0 } \phi _0 (p) \sigma (-p) \, \right\} \; .
\nonumber
\ena
Note that $S_{eff}^{(2)}$ depends only on $M_0$ and $M$, which
are invariant under $SU(2)$ rotations. This is the desired result,
since the average over the $SU(2)$ orientations can now be
trivially performed.

We are now going to study the occurrence of an imaginary part
of the scalar correlation function signaling a quark-anti-quark
threshold. The crucial observation will be that whenever the trace
of the color indices is required, the contributions with
$\tilde{M}^{\alpha }$ occur in conjugate complex pairs. In particular,
this will erase imaginary parts of the correlations function, and no
quark-anti-quark threshold will occur.

In order to work out this phenomena in some detail for the scalar correlation
function (\ref{eq:9}), it is sufficient to study the functions
$H_0(p^2)$, (\ref{eq:e8}) and $H_v(p^2)$, (\ref{eq:e9}), since
they provide the complete correlator with the help of (\ref{eq:e12}).
We rewrite, for instance, $H_0(p^2)$ as
\be
H _0(p^2) \; = \; \frac{1}{4 \pi ^2}  \int _0^1 d \alpha
\int_0^{\Lambda ^2} du  \; \left\{ 1 \, - \, \frac{3Q}{ u + Q }
\, + \, \frac{2 Q^2 }{ [u + Q ]^2 } \right\}
\; + \;  (M \rightarrow -M) \; .
\label{eq:14}
\en
To illustrate the disappearance of the quark-anti-quark threshold
for $M \not= 0$, we first study its occurrence for $M=0$. In this case
our model describes the scalar correlation function in the usual
constituent quark model with a constituent quark mass $M_0$. The term
of interest is the second one in the curly bracket in (\ref{eq:14}).
After integration over $u$, this term becomes essentially $\ln Q$.
This implies that whenever $Q$ becomes negative, the function $H_0$,
(\ref{eq:14}), acquires an imaginary part. In order for $Q$ to become negative,
the Euclidean momentum $p^2$ must satisfy,
\be
- p^2 \; < \; 4 M_0^2 \; ,
\label{eq:16}
\en
implying that the quark-anti-quark threshold occurs at a
(Minkowskian) momentum
$p_M = 2 M_0$, which is the familiar result. For $M=0$ the function
$H_v(p^2)$ does no harm, since it is identically zero by
definition, (\ref{eq:e9}).

We are now going to show that for $M \not= 0$ no threshold will occur at all.
Adding up the contributions from $M$ and $-M$ in (\ref{eq:14}), the
crucial term becomes
\be
H_0^{crit} \; = \; - \frac{3}{2 \pi ^2}
\int _0^1 d \alpha \int_0^{\Lambda ^2} du \;
\frac{ (u + W)W + 4 M_0^2 M^2 }{ [ u + W ]^2 + 4 M_0^2 M^2 } \; ,
\label{eq:17}
\en
where $W = \alpha (1-\alpha) p^2 + M_0^2 - M^2$. An analogous result
holds for the function $H_v(p^2)$. One finds
\be
H_v^{crit} \; = \; - \frac{3}{2 \pi ^2}
\int _0^1 d \alpha \int_0^{\Lambda ^2} du \;
\frac{ 2 M_0^2 M^2 }{ [ u + W ]^2 + 4 M_0^2 M^2 } \; .
\label{eq:17a}
\en
We find that the logarithmic divergence is screened if $M M_0 \not=0$.
No imaginary part occurs for $M_0 \not= 0$ {\it and } $M \not= 0$.
This is our main observation. For non-vanishing current mass $m$,
one always expects  at least a small constituent quark mass $M_0$.
Therefore, the main ingredient in avoiding the
decay of the scalar meson into a quark-anti-quark pair is the
non-vanishing value of $M$. In the following, we refer to the phase
of the constituent quark model (\ref{eq:1}) with $M \not=0 $
as {\it confining phase}.

In the chiral limit, the chiral symmetric phase
$(M_0=0)$ needs further discussion. In section (\ref{sec:4}) we will
find that temperature induces a deconfinement phase transition, and
that chiral symmetry is restored at the same time. Here we find
by an inspection
of (\ref{eq:17},\ref{eq:17a}) that the restoration of
chiral symmetry $(M_0=0)$ is accompanied by the occurrence of
quark-anti-quark thresholds.

\subsection{ Phase structure }
\label{sec:3.2}
\bs

Here we will search for solutions of the gap equations (\ref{eq:7}-\ref{eq:8})
with a non-vanishing, imaginary constituent quark mass in the color-triplet
channel, which implies the remarkable consequences discussed
above. We will discuss the dependence of the phase structure, and,
in particular,
the phase transition from the confining phase $(M\not=0)$ to a non-confining
phase $(M=0)$, on the parameters of the model, i.e. $G_0$,
$G^{\alpha \beta } $ and $m$. For this purpose, we have to analyze
the solution of the gap equations (\ref{eq:7}-\ref{eq:8}).
In order to solve (\ref{eq:8}), we assume $M^{\alpha }$ to be an
eigenvector of the matrix $G^{\alpha \beta }$, i.e.
\be
G^{\alpha \beta } M^{\beta } \; = \; G_c \, M^{\alpha } \; .
\label{eq:18}
\en
This reduces eqs. (\ref{eq:7}) and (\ref{eq:8}) to
(see Appendix \ref{app:b})
\bea
\frac{1}{G_0} (M_0 - m) &=& \; -  M_0 \, I_0(M_0,M) \; + \;
M \, I_v(M_0,M) \; ,
\label{eq:19} \\
\frac{1}{G_c} M &=& - M \, I_0(M_0,M) \; - \; M_0 \, I_v(M_0,M) \; ,
\label{eq:19a}
\ena
where
\bea
I_{0}(M_0,M) &=& - \frac{1}{2 \pi ^2 } \int _0 ^{\Lambda ^2 }
du \; u \;
\frac{ u + M_0^2 - M^2 }{ u^2 + 2 u (M_0^2-M^2) + (M_0^2 + M^2 )^2} \; ,
\label{eq:20} \\
I_{v}(M_0,M) &=&  \frac{1}{ \pi ^2 } \int _0 ^{\Lambda ^2 }
du \; u \;
\frac{ M_0 M }{ u^2 + 2 u (M_0^2-M^2) + (M_0^2 + M^2 )^2 } \; ,
\label{eq:20a}
\ena
where a sharp O(4) cutoff $\Lambda $ was introduced.
Since $I_v$ is proportional to $M_0 M$, $M=0$ is always a solution of
eq. (\ref{eq:19a}). In this case, eq. (\ref{eq:19})
is reduced to the gap equation of the standard Nambu-Jona-Lasinio model
with an additional $SU(2)$ degree of freedom. This implies that the standard
Nambu-Jona-Lasinio model is contained in the extended model (\ref{eq:1})
as a special case, and we expect the theory (\ref{eq:1}) to be
phenomenologically
as successful as the Nambu-Jona-Lasinio model.

The solution of the system of eqs. (\ref{eq:19},\ref{eq:19a})
was investigated numerically.
Figure \ref{fig:1} shows the color-singlet and color-triplet
constituent quark masses, $M_0$ and $M$,  as function of the color-singlet
coupling $G_0$. For sufficiently small color-singlet coupling strength
a confining phase with $M \not= 0$ exists. Also shown is the non-confining
phase with $M=0$ for all values of $G_0$. In order to decide which
solution forms the vacuum, one has to compare the classical action, i.e.
\be
{\cal A}_c = \int \frac{ d^{4}k }{ (2 \pi )^{4} } \; \left\{
-2 \ln [ ( k^2 - M_0^2-M^2)^2+4 M_0^2 k^2 ] \, - \,
\frac{1}{2 G_c} M^2 \, + \, \frac{1}{2 G_0} (M_0 -m )^2 \right\} ,
\label{eq:21}
\en
of both solutions. One finds that the confining solution has a lower
action and forms the vacuum.

It is observed that the color-triplet coupling strength $G_c$
must exceed a critical value in order to allow for the desired imaginary
constituent quark mass in the color-triplet channel. Figure
\ref{fig:2} shows the critical coupling as function of the
color-singlet coupling $G_0$ for different values of the current mass $m$.
A strong color-singlet constituent mass $M_0$ seems to suppress
the occurrence of the confining phase. The situation can be compared with
that of free electrons in a solid. It was discovered by Anderson that the
electrons get localized if the density of impurities exceeds a critical
limit~\cite{anderson}.

Finally, we present the result for the scalar correlation function
$\Delta _s^c (p^2)$, (\ref{eq:10})
which we have calculated in the last subsection.
The final result (two integrations are left to a numerical calculation)
is given by
\be
\Delta _c^s (p^2) \; = \; \frac{1}{G_0} \, - \, \frac{1}{G_0^2}
\frac{1}{ \frac{1}{G_0} - H_0 (p^2) \, + \,
\frac{ H_v^2(p^2) }{ \frac{1}{G_c} - H_0(p^2) } } \; .
\label{eq:21a}
\en
Figure \ref{fig:3} shows the correlation function $\Delta _s^c(p^2)$
as function of the Euclidean momentum transfer $p^2$ for
different values for the current quark mass $m$. From Figure
\ref{fig:2}, we conclude that increasing the current mass $m$
at fixed values of $G_0$ and $G_c$ drives the system towards the
deconfinement phase transition. For large $m$ (close to the critical
value of $m$ where the transition to the
deconfined phase occurs) one observes a resonance-like peak
at a negative momentum squared, which is reminiscent of the quark
anti-quark threshold of the deconfined phase ($M=0$).
In the latter case, our model
is identical to the Nambu-Jona-Lasinio model. It is well known that
in the standard Nambu-Jona-Lasinio model, a weakly bound scalar particle
occurs in the spectrum. It has a mass of two times the constituent quark
mass implying that the particle pole occurs at the threshold position.
The physics of the scalar meson is therefore beyond the scope of the
standard Nambu-Jona-Lasinio model.
This defect of the NJL-model is cured in our model (\ref{eq:1}):
the threshold is avoided in the confining phase $(M\not=0)$. At the
same time, no scalar particle is present in the spectrum. There appears
a ``resonance" near the threshold position
with a width proportional to $M$. This result is in agreement with
nature, since no scalar bound state is observed in the meson
spectrum at the relevant energy scale.
The scalar Green's function
develops a structure around the threshold position without allowing
the meson to decay into two quarks. This might be a
precursor to a parton structure of hadrons at high energy. Unfortunately
we cannot push this conjecture further, since our model is limited
to low energies.

One might raise the question as to whether this result is in not disagreement
with dispersion relations, since the correlation function we have is a real
function in the entire momentum space. It is interesting to note
that our model does {\it not} allow a scalar particle to appear,
although a scalar field is needed in the bosonization approach to
represent the quark interaction. The scalar correlation function
at hand is, therefore, not appropriate for discussing dispersion relations.
On the other hand,
as we will see soon, the pion-pion correlation function exhibits
the pion pole and is otherwise real, so one might worry about
the dispersion relation in this case. Note, however, that in order to
arrive at this result, we have expanded the effective meson theory
(\ref{eq:5}) up to second order in the meson fields ignoring
mesonic interactions. If these interactions are included, we expect to recover
the physical thresholds of an interacting meson theory
satisfying dispersion relations.

Although the correlation function is real, a non-trivial structure
appears at the would-be threshold position. A close inspection of
the correlator (\ref{eq:21a}) (see Appendix \ref{app:b2}) shows that
the function possesses cuts in the complex $p$-plane. One might then
question whether these complex structures violate fundamental
principles of quantum field theory (e.g.\ stability). We note that
our approach -- the bosonization procedure of section \ref{sec:2.2} and
the related approximations -- is supposed not to violate any of the
analyticity requirements. In comparison to other NJL-type
models, our ground state (vacuum) is simply more complicated, with the standard
NJL model corresponding to a particular case ($M=0$).
It may be that some of the fundamental requirements of QCD
are not correctly implemented. If so, then
our detailed calculation may be providing mechanisms for the cancellation of
unwanted features. In this case, we could expect our model to give some
insight in the analyticity structure of Green's functions of a
confining theory. This question is of particular interest, but is
beyond the scope of the present paper. See \cite{oehme} for a discussion
on this matter. To check that nothing is amiss with our model,
we have studied the analytic structure
of the scalar correlation function in Appendix \ref{app:b2} and
explicitly verified that the stability criterion is indeed satisfied thanks to
a cancellation mechanism mentioned above.

\newpage
\section{ Chiral Properties }
\label{sec:ch}
\bs

In this section we study the properties of the pion that
should emerge as a Goldstone boson of the spontaneously broken
chiral symmetry. We will show that our model is in agreement with
the low-energy theorems, and will verify by an explicit calculation that
the Gell-Mann-Oakes-Renner relation holds.

\subsection{The pion Bethe-Salpeter equation }
\label{sec:ch1}
\bs

One way to extract the properties of the pion is to
study the pseudoscalar correlation function in the
color-singlet channel which can be obtained in the same way
as for the scalar correlator discussed in section \ref{sec:3.1}.
Here we prefer to use a different approach which is to calculate the
pion Bethe-Salpeter amplitude. This method will illustrate
the role of the hidden color components of the pion, and
will readily provide an access to such observables as the pion decay
constant and the pion electromagnetic form factor.

The Bethe-Salpeter equation for the pion amplitude
$(P_0, P^\alpha )$ is directly obtained from the
effective meson Lagrangian ${\cal L}_M$ (\ref{eq:5}) by
\be
\sum _{b=0,1\ldots 3} \;
\frac{ \delta ^2 {\cal L}_M }{ \delta \pi ^{a}(-p)
\delta \pi ^{b} (p) } \, P^b(p) \vert_ {p^2= - m_\pi ^2} \; = \; 0 \; .
\label{eq:h1}
\en
For a fixed orientation of the interaction matrix $G^{\alpha \beta }$,
the left-hand side of this equation becomes
\be
\left( \begin{array}{cc}
 \frac{1}{G_0} + \Tr \{ \gamma _5 S(k+p) \gamma _5 S(k) \} &
 \Tr \{ \tau ^\alpha \gamma _5 S(k+p) \gamma _5 S(k) \} \cr
 \Tr \{ \tau^ \alpha \gamma _5 S(k+p) \gamma _5 S(k) \} &
 (G^{-1})^{\alpha \beta } + \Tr \{ \tau ^\alpha \gamma _5 S(k+p)
 \tau ^\beta \gamma _5 S(k) \} \cr
       \end{array} \right) \;
\left( \begin{array}{c}
 P_0 \cr P^{\alpha } \cr
       \end{array} \right) ,
\label{eq:h2}
\en
where the trace $\Tr $ extends over the momentum space ($k$) as well as
Lorentz- and color-space. One observes that the ansatz
\be
P^\alpha \; = \; i \frac{ M^{\alpha } }{M} P_1
\label{eq:h3}
\en
for the color-components of the pion\footnote{
Once the average over all orientations of the interaction matrix
is performed, these parts become the hidden-color components
(compare section \ref{sec:ch3}).}
reduces (\ref{eq:h2}) to (see Appendix \ref{app:c})
\be
\left( \begin{array}{cc}
 \frac{1}{G_0} + {\cal I }_0(p^2)  &
 - {\cal I}_v(p^2) \cr
 {\cal I}_v(p^2) & \frac{1}{G_c} + {\cal I}_0(p^2) \cr
       \end{array} \right) \;
\left( \begin{array}{c}
 P_0(p^2) \cr P_1(p^2) \cr
       \end{array} \right) \; = \; 0 \; ,
\label{eq:h4}
\en
where the functions ${\cal I}_0$ and ${\cal I}_v$ are defined by
\bea
{\cal I}_0 (p^2) &=& - \frac{1}{4 \pi ^2} \int _0^1 d\alpha \;
\int _0 ^{\Lambda ^2 } du \; u \;
\frac{ u - \alpha (1- \alpha ) p^2 + A^2 }{
\left[ u + \alpha (1- \alpha ) p^2 + A^2 \right] ^2 }
\; + \; (M \rightarrow -M) \; ,
\label{eq:h5} \\
{\cal I}_v (p^2) &=&  \frac{i}{4 \pi ^2} \int _0^1 d\alpha \;
\int _0 ^{\Lambda ^2 } du \; u \;
\frac{ u - \alpha (1- \alpha ) p^2 + A^2 }{
\left[ u + \alpha (1- \alpha ) p^2 + A^2 \right] ^2 }
\; - \; (M \rightarrow -M) \; ,
\label{eq:h6}
\ena
with $A=M_0+iM$.
An inspection of (\ref{eq:h6}) yields that both
${\cal I}_0$ and ${\cal I}_v$ are real. Demanding
({\ref{eq:h4}) to have a non-trivial solution for $P^2=-m_\pi ^2$
leads to  a nonlinear equation to determine $m_\pi $. Instead of
solving this equation numerically, it is more instructive to solve it
analytically for pion masses small compared with the constituent quark mass
$M_0$.
For this purpose we expand the functions ${\cal I}_0$ and ${\cal I}_v$
up to linear order in $p^2$, i.e.
\be
{\cal I}_0(p^2) \; = \; I_0 \, + \, F_0 \, p^2 \, + O(p^4) \; , \hbo
{\cal I}_v(p^2) \; = \; I_v \, + \, F_v \, p^2 \, + O(p^4) \; ,
\label{eq:h7}
\en
where $I_0$ and $I_v$ depend, respectively,  on $M_0$ and $M$ (see
(\ref{eq:20}) and {\ref{eq:20a})), and the quantities
$F_0$, $F_v$ are defined by
\bea
F_0 &=& \frac{1}{8 \pi ^2} \int _0^{\Lambda ^2} d(k^2) \; k^2 \;
\frac{1}{ \left[ k^2 + (M_0+iM)^2 \right] ^2 }
\; + \; (M \rightarrow -M) \; ,
\label{eq:h8} \\
F_v &=& - \frac{i}{8 \pi ^2} \int _0^{\Lambda ^2} d(k^2) \; k^2 \;
\frac{1}{ \left[ k^2 + (M_0+iM)^2 \right] ^2}
\; - \; (M \rightarrow -M) \; ,
\label{eq:h9}
\ena
A non-trivial solution for the Bethe-Salpeter amplitudes $P_0$,
$P_1$ exists provided
\be
\det \left( \begin{array}{cc}
 \frac{1}{G_0} + I_0 + F_0 p^2  &
 - I_v - F_v p^2 \cr
 I_v + F_v p^2 & \frac{1}{G_c} + I_0 + F_0 p^2 \cr
       \end{array} \right) \; = \; 0 \; .
\label{eq:h10}
\en
With the help of the equations of motion (\ref{eq:19}) and
(\ref{eq:19a}), i.e.
\be
\frac{1}{G_0} + I_0 \; = \; \frac{M}{M_0} I_v \, + \,
\frac{m}{M_0 G_0} \; , \hbo
\frac{1}{G_c} + I_0 \; = \; - \frac{M_0}{M} I_v \; ,
\label{eq:h11}
\en
eq. (\ref{eq:h10}) can be rewritten as (setting $p^2=-m_\pi ^2$)
\be
m_\pi ^2 \, f_{\pi c} ^2 \; = \; 4m \; \langle \bar{q} q \rangle \; + \;
O(m_\pi^4) \, + \, O(m^2) \; ,
\label{eq:h12}
\en
where we have used that the quark condensate $\langle \bar{q}q \rangle $
is given by $M_0 / G_0$, and where $f_{\pi c}^2$ is defined by
\be
f_{\pi c}^2 \; = \; 4 \left\{ M_0^2 F_0 \; - \;
M^2 F_0 \; - \; 2 M M_0 F_v \right\} \; .
\label{eq:h13}
\en
Equation (\ref{eq:h12}) is the Gell-Mann-Oakes-Renner relation.
It tells us that in the chiral limt $(m=0)$ the pion is massless
if chiral symmetry is spontaneously broken $(\langle \bar{q}q \rangle
\not=0)$. Below, we will show that $f_{\pi c}^2$, defined
in (\ref{eq:h13}), coincides with the pion decay constant
in the chiral limit $(m=0)$.

\subsection{ The electromagnetic form factor }
\label{sec:ch2}
\bs

Since the Bethe-Salpeter equation provides only the relative
weights of the amplitudes $P_0$, $P^\alpha $, an additional
physical input is needed for the normalization of the amplitudes.
One may use the electromagnetic form factor of the pion.  The
electromagnetic form factor provides the desired additional
constraint. The form factor $F(q^2)$ is defined via the electromagnetic
vertex~\cite{goe83}, i.e.
\be
\langle \pi (p') \vert J_\mu (0) \vert \pi (p) \rangle \; = \;
ie \, (p'_\mu + p_\mu) \, F(q^2) \; ,
\label{eq:h14}
\en
where $J_\mu $ is the electromagnetic current and $q=p'-p$.
In order to normalize the pion charge to unity, we demand that
\be
F(0) \; = \; 1 \; .
\label{eq:h15}
\en
Once the form factor is calculated, the (mean-square)
charge radius of the pion is simply given by
\be
r_{ms}^2 \; = \; \langle r^2 \rangle _{\pi } \; = \;
- 6 \, \frac{ \partial F(q^2) }{ \partial q^2 } \vert _{q^2=0} \; ,
\label{eq:h16}
\en
where $q$ is the momentum in Minkowski space.
Writing the electromagnetic current
in terms of the quark fields,
\be
J_\mu (x) \; = \; \bar{q}(x) \, \gamma _\mu \, q(x) \; ,
\label{eq:h17}
\en
we can study the matrix element (\ref{eq:h14}) most efficiently by using
the Bethe-Salpeter amplitude of the pion. Its
graphical representation is given in Fig.\ref{fig:ch1}\footnote{
Since the diagrams (in particular, (a)) are divergent, the result depends
on the actual choice of the loop momentum. A more sophisticated
cutoff procedure (e.g.\ Schwinger's proper-time regularization) would
remove this ambiguity. }.
For $p'=0$, the matrix element is
\bea
- \int \frac{ d^4k }{ (2\pi )^4 } \; \tr \bigg\{ &\gamma _5&
\left( P_0 + i \frac{ M^\alpha }{M} \tau ^\alpha P_1 \right) \,
S(k- \frac{p}{2} ) \, \gamma _\mu \, S(k+ \frac{p}{2} ) \,
\label{eq:h18} \\
&\gamma _5&
\left( P_0 + i \frac{ M^\beta }{M} \tau ^\beta P_1 \right) \,
S(k- \frac{p}{2} ) \, \bigg\} \; .
\nonumber
\ena
The evaluation of this matrix element is left to Appendix \ref{app:d}.
With (\ref{eq:d7}), the form factor, for small momentum, reads
\be
F(p^2) \; = \; \left( P_0^2 - P_1^2 \right) \, \left(
F_0 - p^2 R_0 \right) \; - \; 2 P_0 P_1 \, \left(
F_v - p^2 R_v \right) \; .
\label{eq:h19}
\en
{}From this expression follow the desired normalization
(\ref{eq:h15}) of
the pion Bethe-Salpeter amplitude and the pion charge radius
(\ref{eq:h16}):
\bea
1 &=& \left( P_0^2 - P_1^2 \right) \, F_0 \; - \; 2 P_0 P_1 \, F_v  \; ,
\label{eq:h20} \\
r_{ms}^2 &=& 6 \left[ \left( P_0^2 - P_1^2 \right) \, R_0 \; - \;
2 P_0 P_1 \, R_v  \, \right] \; .
\label{eq:h21}
\ena

We can now discuss the full momentum dependence of the form factor
$F(p^2)$. The expression for $F(p^2)$ is derived in Appendix
\ref{app:d}. The integration over the angle variable and the
radial component of the momentum space was performed numerically.
The numerical result is presented in Figure \ref{fig:ch3}.
There exists no vector-meson poles at negative momentum transfer, but
instead a peak, its width strongly depending on the parameters of
the model. Presumably
the resonance-like structure is an artifact of the
model, unrelated to any physical processes. It probably reflects
the suppression of the quark-anti-quark pole present in the
standard NJL model without confinement.  As mentioned, the physics
of the vector mesons is beyond the scope of a local four-quark
interaction\footnote{ In particular, we do not have any vector and
axial-vector channels in the quark interaction (\ref{eq:2}),
which are known to be important for the physics of vector mesons. }.
We expect that a more realistic model with quantum loops
will produce a vector-meson (i.e., $\rho$) pole
instead of the resonance. In fact, we observe a pion pole and no further
structure in the pseudoscalar correlation function in our model,
which is supposed to give good results for the pion physics (see below).

It is quite remarkable that we obtain a real
form factor for arbitrary large (Minkowskian)
momentum
manifesting the absence of quark and anti-quark thresholds.
This cures an outstanding problem of the standard non-confining NJL-model.
This may have an important consequence in heavy-meson physics.
For instance, the non-perturbative description of the electroweak currents
of a heavy and a light quark is beyond the reach of the usual NJL-model
due to the occurrence of quark anti-quark threshold effects.
Electroweak currents such as the one considered here enter into the decay rate
of a B-meson into a pion and an electron. In the limit of the
electron being very fast, one should be able to extract the element
$V_{bu}$ of the Kobayashi-Maskawa mixing matrix from the
differential cross section. However, it has been shown that
in the limit of a fast electron, non-perturbative
contributions to the electroweak currents become important.
We believe that a more realistic formulation ($SU(3)$ color)
of our model might be able to provide the desired results.

The normalization (\ref{eq:h20}) completes the calculation of the
pion Bethe-Salpeter amplitude. This is the desired result, since
we are now able to calculate matrix elments involving pions.
In particular, we will obtain the pion decay constant in the
next chapter by a direct calculation.

\subsection{ The pion decay constant }
\label{sec:ch3}
\bs

An expression of pion decay constant was already deduced in
section \ref{sec:ch1} by demanding that the Gell-Mann-Renner-Oakes
relation (\ref{eq:h12}) be quantitatively satisfied. The
aim of this subsection is to show that the decay constant (\ref{eq:h13}),
obtained there, is identical to that derived from its definition.

The pion decay constant $f_\pi $ is defined by the matrix element
\be
\langle 0 \vert A_{\mu }(x) \vert \pi (p) \rangle
\vert _{p \rightarrow 0} \; = \;
i \, e^{ipx} \, p_\mu \, f_\pi  \; ,
\label{eq:h22}
\en
which describes the coupling of the pion to the vacuum via the
axial vector current $A_{\mu }(x) = \bar{q}(x) \gamma _5 \gamma _\mu
q(x) $. The matrix element (\ref{eq:h22}), also shown in
Figure \ref{fig:ch1}, is
\be
- \int \frac{ d^4k }{ (2\pi )^4 } \; \tr \left\{
\gamma _5 \gamma _\mu \, S(k+p) \, \left( P_0 + i \frac{ M^{\alpha } }{M}
P_1 \right) \gamma _5 \, S(k) \, \right\} \; .
\label{eq:h23}
\en
An average over all color orientations of the interaction matrix
$G^{\alpha \beta }$ -- which is equivalent to an average over all
color directions of $M^\alpha $ --  is understood in (\ref{eq:h23}).
A straightforward calculation of (\ref{eq:h23}) for a given
vector $M^\alpha $ yields
\bea
4i \, p_\mu &\bigg\{ &  \int \frac{ d^4k }{ (2\pi )^4 } \;
\left[ \frac{ A }{ \left[ k^2 + A^2 \right] ^2 } + (M \rightarrow -M)
\right] \; P_0
\label{eq:h24} \\
&-& (-i) \int \frac{ d^4k }{ (2\pi )^4 } \;
\left[ \frac{ A }{ \left[ k^2 + A^2 \right] ^2 } - (M \rightarrow -M)
\right] \; P_1 \; \bigg\} \; .
\nonumber
\ena
This result can be further simplified by introducing the functions
$F_0$ and $F_v$ from (\ref{eq:h8},\ref{eq:h9}), i.e.
\be
2i \, p_\mu \, \left\{
(M_0 F_0 - M F_v) \, P_0 \; - \; (M_0 F_v + M F_0) \, P_1
\right\} \; .
\label{eq:h26}
\en
The ratio $\epsilon $ of $P_0$ over $P_1$ is provided by the
Bethe-Salpeter equation (\ref{eq:h4}), i.e.
\be
\epsilon \; := \; \frac{ P_1 }{ P_0 } \; = \; - \frac{
{\cal I}_v(p^2=-m_\pi ^2) }{ \frac{1}{G_c} +
{\cal I}_0(p^2=-m_\pi ^2) } \; ,
\label{eq:h25a}
\en
whereas the overall normalization is constrained by the
electromagnetic form factor (\ref{eq:h20}). Expressing $P_1$
in terms of $\epsilon $ and $P_0$ and eliminating $P_0$ with the
constraint (\ref{eq:h20}), we obtain the final result
\be
f_\pi ^2 \; = \; 4 \, \frac{ \left[
M_0 F_0 - M F_v \; - \; (M_0 F_v + M F_0) \, \epsilon
\right] ^2 }{ (1 - \epsilon ^2) \, F_0
\, - \, 2 \epsilon F_v } \; ,
\label{eq:h25}
\en
where the functions $F_0$, $F_v$ are given by (\ref{eq:h8}) and
(\ref{eq:h9}).
This is our result for the pion decay constant, valid for all values
of the current mass. This decay constant needs to agree with the one
extracted from the Gell-Mann-Renner-Oakes relation, (\ref{eq:h13}),
only in the chiral limit $(m=0)$. Before we show that this is
indeed the case, we would like to comment on the result (\ref{eq:h25}).
First note that the average of all orientations has still to be
performed. Since our result (\ref{eq:h25}), however, depends only
on the invariant quantity $M$ rather than on $M^\alpha $, this average
is trivial. Note further that a term proportional to $P_1$ enters into
the physical decay constant. This is the contribution from the
hidden-color components of the pion.

In order to calculate $f_\pi $ in the chiral limit, we proceed with
(\ref{eq:h24}).
{}From the Bethe-Salpeter equation (\ref{eq:h4}) in the chiral limit
$p^2 = - m_\pi ^2 = 0$, one finds
\be
P_0 \; = \; \frac{ M_0 }{ M } \, P_1 \; .
\label{eq:h27}
\en
With this result, the decay constant in the chiral limit becomes
\be
f_{\pi c} \; = \; 2 \left\{
M_0^2 F_0 \, - \, 2 M M_0 F_v \, - \, M^2 F_0 \right\} \;
\frac{ P_1 }{M} \; .
\label{eq:h28}
\en
Using (\ref{eq:h27}) in the normalization of the
Bethe-Salpeter amplitude (\ref{eq:h20}), one may
eliminate $P_1$ in (\ref{eq:h28}) to obtain
\be
f_{\pi c}^2 \; = \; 4 \left\{ (M_0^2 -M^2) \, F_0 \; - \; 2 M M_0
\, F_v \right\} \; .
\label{eq:h29}
\en
This expression is identical to (\ref{eq:h13}). This
establishes that the Gell-Mann-Oakes-Renner relation is
indeed valid in our model.

We finally present the numerical data for the pion mass and the
pion decay constant. We have numerically solved the
Bethe-Salpeter equation (\ref{eq:h4}) to obtain the mass of the pion
and the ratio $\epsilon = P_1/P_0$. The decay constant $f_\pi $
was then calculated from (\ref{eq:h25}). The result for
$m_\pi $ and $f_\pi $ as a function of the current mass is shown in
Figure \ref{fig:ch2}. One observes that $m_\pi ^2$ depends almost linearly
on the current mass with the slope given by the Gell-Mann-Oakes-Renner
relation. The decay constant $f_\pi $ decreases with
increasing current mass.

\subsection{ The pseudoscalar correlation function }
\label{sec:ch4}
\bs

It is instructive to compare the result of the scalar correlation
function with that for the pseudoscalar correlator where one
expects a pion pole. The pseudoscalar
correlation function is defined by
\be
\Delta _\pi ^c(p) \; = \; \int d^{4}x \; e^{-ipx} \;
\frac{ \delta ^2 \, \ln Z[\phi ] }{ \delta \phi _5 (x)
\delta \phi _5(0) } \vert _{\phi =0 } \; .
\label{eq:p1}
\en
In order to derive $\Delta _\pi ^c$, we expand the effective meson
theory (\ref{eq:4}) up to second order in pion fields.
Since an average over all orientations of the interaction matrix
$G^{\alpha \beta }$ is required, it is convenient to integrate out
the colored pion fields. One observes that the resulting
theory for the color-singlet pion does not depend
on this orientation, so the averaging is trivial. Since the
explicit calculation parallels closely that of the
scalar correlator in section \ref{sec:3.1}, we shall simply present
the final result:
\be
S_{eff}^{(2)} \; = \; \int \frac{ d^4p }{ (2\pi )^4 } \; \left\{
\frac{1}{2} \pi (p) \, \Pi _5 (p^2) \, \pi (-p)
\; + \; \frac{1}{ 2 G_0 } \phi _5 (p) \phi _5(-p)
+ \frac{1}{ G_0 } \phi _5 (p) \pi (-p) \, \right\} \; ,
\label{eq:p2} \\
\en
where the pion dispersion formula $\Pi _5 (p^2)$ is given by
\be
\Pi _5 (p^2) \; = \;
\frac{1}{G_0} + {\cal I}_0 (p^2) \, + \,
\frac{ {\cal I}_v^2(p^2) }{ \frac{1}{G_c} + {\cal I}_0(p^2) } \; .
\label{eq:p4}
\en
The functions ${\cal I}_0$ and ${\cal I}_v$ were defined in
(\ref{eq:h5}) and (\ref{eq:h6}), respectively.
The dispersion formula $\Pi _5$ is shown in Figure \ref{fig:3b}
as function of the Euclidean momentum transfer in the chiral limit
$(m=0)$. Note that the function $\Pi _5 (p^2)$ is zero at
$p^2=0$. This zero gives rise to a pole of the correlation
function at zero momentum transfer confirming the pion as
Goldstone boson. The most striking feature of Figure \ref{fig:3b}
is a singularity at positive Euclidean momentum. It arises
from the second term in (\ref{eq:p4}), which stems from the
integration over the colored pions. The singularity occurs at the
momentum where the colored pions go on-shell. The occurrence
of the singularity therefore is quite natural and does not
depend on the details of the model. In our model, the colored
pions go on-shell at positive Euclidean momentum $p^2$, which
might indicate that the colored pions condense. In this case
our approach, expanding the Lagrangian up to second order in the
colored pion fields and integrating them out, is no longer
appropriate. We will leave this issue to future investigations
and here study the pion pole in some detail.

One can deduce the pion mass $m_\pi^2$
from the position of the pion pole in the correlation function
$\Delta _\pi ^c (p^2)$.
It is worthwhile to check whether this result for the
pion mass agrees with that obtained from solving the
Bethe-Salpeter equation (see section \ref{sec:ch2}). One observes
that both masses are indeed identical, because the constraint for the
dispersion relation to be satisfied coincides with the
condition (\ref{eq:h10}) which guarantees a non-zero
Bethe-Salpeter amplitude. Finally the numerical result
for the correlator (\ref{eq:p1}) is presented in Figure \ref{fig:3a}.
The correlation function is entirely dominated by the pion pole.
This is compared with the result for the scalar correlator.
In the latter case, no pole occurs at all, but the scalar correlation
function is significantly influenced by a peak structure.

A second pole in the correlation function occurs at a positive
value of the Euclidean momentum $p^2$. This pole is due to the
fact that the occurrence of the colored pion pole in the
dispersion relation $\Pi _5 $ gives rise to a further zero
of $\Pi _5 (p^2)$. The pole of $\Delta _\pi ^c$ at a positive
momentum squared is intimately related to the properties of the
hidden-color pions and will be the subject of a future work.

\section{ Temperature Effects }
\label{sec:4}
\bs

The model (\ref{eq:1}) offers a confining phase and a deconfining
phase, both emerging from the classical equations of motion
(\ref{eq:7}-\ref{eq:8}). In the section given above, we
investigated the phase structure in the model's parameter space.
We found that the ground state is in the confining phase
if the color-triplet coupling strength is strong enough.
In this section, we will investigate the influence of temperature
on the phase structure. In particular, we will start from the system
with the vacuum in the confining phase and will study the
type of phase transitions, if any, to the deconfining phase.

For this purpose, we must generalize the gap equations
(\ref{eq:7}-\ref{eq:8}) to finite temperature. The trace in the first
terms on the left-hand side stems from the integration over the quark
fields. In order to introduce temperature, we adopt the usual
imaginary-time formalism~\cite{ka89} and
confine the configuration space of the fermionic
fields to the configurations which are anti-periodic in the Euclidean time
direction with a periodic length $1/T$ with $T$ the temperature.
At finite temperature, the integration over the zeroth component of the
Euclidean momentum in the terms of (\ref{eq:7}-\ref{eq:8}) is
replaced by a discrete sum over Matsubara frequences. In order to
illustrate the evaluation of such trace terms at finite temperature,
we explicitly work out a term, which arises in (\ref{eq:7}-\ref{eq:8})
once the color trace is performed, i.e.
\be
F \; := \; \Tr \left\{ \frac{i}{ \kslash + i(M_0 + iM) } \right\} \; .
\label{eq:22}
\en
Performing the Dirac trace as well as the trace over space-time,
one obtains
\be
F \; = \; 4V \sum _{n = -\infty }^{\infty } \int
\frac{ d^{3}k }{(2\pi )^3 } \;
\frac{ M_0 + i M }{ \pi ^2 T^2 (2n+1)^2 + (a + ib)^2 } \; ,
\label{eq:23}
\en
where $V$ is the space volume, and the quantities
\bea
a &=& \sqrt{ \frac{1}{2} \left[ \vec{k}^2 + M_0^2 - M^2 +
\sqrt{ (\vec{k}^2 + M_0^2 - M^2 )^2 + 4 M_0^2 M^2 } \right] }
\label{eq:24} \\
b &=& \frac{ M_0 M }{a}
\label{eq:25}
\ena
were introduced as an abbreviation. In order to split the
zero temperature part, which is cutoff-dependent, from the
temperature-dependent contributions, which are finite,
we perform a Poisson resummation
of the Matsubara sum in (\ref{eq:23}), i.e.
\be
F \; = \; 4 V \int dn \; \frac{ d^{3}k }{(2\pi )^3 } \;
\frac{ M_0 + i M }{ \pi ^2 T^2 (2n+1)^2 + \vec{k}^2 +
(M_0 + iM)^2 } \; + \; F(\nu \not=0) \; .
\label{eq:26}
\en
The first term is the term with zero conjugate Matsubara frequency
($\nu =0$), which yields the zero temperature result if we
substitute $k_0 = \pi T (2n+1)$ for $n$ in the integral in (\ref{eq:26}).
It is this term which needs regularization. It will be performed by introducing
a sharp O(4) cutoff as it was done in the zero-temperature case.
The part of the non-zero conjugate Matsubara indices is finite and
needs no regularization. It is
\be
F(\nu \not=0) \; = \; - \frac{4 V}{T} \frac{ M_0 + i M }{ a + ib }
\left[ e ^{ \frac{a + ib }{T} } +1 \right] ^{-1} \; .
\label{eq:27}
\en
Evaluating the trace terms in (\ref{eq:7}-\ref{eq:8}) as sketched above,
one finds that the gap equations acquire an additional term
that contains a temperature dependence. Introducing
\bea
\Sigma _R &=& \frac{ a e^{a/T} \cos (b/T) + a - b e^{a/T} \sin (b/T) }{
(a^2+b^2) ( e^{2a/T} + 2 \cos (b/T) e^{a/t} +1 ) } \; ,
\label{eq:28} \\
\Sigma _I &=& \frac{ a e^{a/T} \sin (b/T) + b + b e^{a/T} \cos (b/T) }{
(a^2+b^2) ( e^{2a/T} + 2 \cos (b/T) e^{a/t} +1 ) } \; ,
\label{eq:29}
\ena
the modified gap equations are
\bea
G_0^{-1} (M_0 - m) &=& M_0 \, I_+(M_0,M) \; - \; \frac{4}{\pi ^2 }
\int dk \; k^2 \; \left( M_0 \Sigma _R + M \Sigma _I \right) \; .
\label{eq:30} \\
G_c^{-1} M &=& M \, I_-(M_0,M) \; - \; \frac{4}{\pi ^2}
\int dk \; k^2 \; \left( M \Sigma _R - M_0 \Sigma _I \right) \; ,
\label{eq:31}
\ena
where the functions $I_{\pm }$ are defined with the help of
(\ref{eq:20},\ref{eq:20a}) by
\be
I_+ \; = \; - I_0 \, + \, \frac{M}{M_0} I_v \; , \hbo
I_- \; = \; - I_0 \, - \, \frac{M_0}{M} I_v \; .
\label{eq:31a}
\en
This is the main result of this section; the
temperature-dependent color-singlet and color-triplet constituent quark
masses will emerge from these equations.

We have studied numerically
the solutions $M$ and $M_0$ of (\ref{eq:30}-\ref{eq:31})
as function of temperature. The coupling strengths
$G_0$ and $G$ were chosen in order for the system to be in the
confining phase at zero temperature.
The result is shown in Figure (\ref{fig:4}). If the temperature exceeds
a critical value, a first order phase transition from the confining
phase $(M \not=0)$ to the deconfined phase $(M=0)$ takes place.
The deconfining phase transition is accompanied by a sudden drop
of the color-singlet constituent mass $M_0$ indicating at the same time the
restoration of chiral symmetry. The small residual constituent quark mass
is due to the current mass $m=0.02 \Lambda $ which explicitly breaks
chiral symmetry.

For phenomenological applications, the dependence of the
deconfining phase transition on the current quark mass is of particular
interest. Figure \ref{fig:5} shows the critical color-triplet
coupling strength $G_c$ as function of the temperature for
some values of the current mass. The result suggests that
quark liberation of {\it all} quark flavors occurs approximately
at the same temperature (assuming that the system was in the confining
phase at zero temperature).


\section{Finite-Density Effects}
\label{sec:5}
\bs

In the last section, we observed a phase transition from the confined
phase to the deconfined phase at high temperature. For phenomenological
reasons~\cite{newbr}, one also expects a phase transition to
occur at high baryonic density. Here we investigate the
existence of this phase transition in the
random background quark model~(\ref{eq:1}).

In order to study the system at a non-zero baryonic density,
we introduce a chemical potential $\mu $ into
the Lagrangian (\ref{eq:2}):
\bea
{\cal L }_D  &=& \bar{q}(x) ( i \dslash + im + i \mu \gamma _0 ) q(x) \; + \;
G_0 [ \bar{q} q(x) \, \bar{q} q(x) \, - \, \bar{q} \gamma _5 q(x) \,
\bar{q} \gamma _5 q(x) ]
\label{eq:40} \\
&+& [ \bar{q} \tau ^{\alpha } q(x)  \, G^{\alpha \beta }
\bar{q} \tau ^{\beta } q(x) \, - \,
\bar{q} \gamma _5 \tau ^{\alpha } q(x)  \, G^{\alpha \beta }
\bar{q} \gamma _5 \tau ^{\beta } q(x)  ] \; ,
\nonumber
\ena
The bosonization procedure -- introducing scalar and pion fields -- is
unchanged by the presence of the chemical potential. The gap equations
(\ref{eq:7},\ref{eq:8}) acquire an additional part due to finite
density, i.e.
\bea
- \frac{1}{V_4} \Tr \left\{ \frac{ i }{ i \dslash + i M_0 + i
\tilde{M}^{\alpha } \tau ^{\alpha } } \right\} &-&
I_F(M_0,M) \; + \; \frac{1}{G_0}
\left( M_0 -m \right) \; = \; 0 \; ,
\label{eq:41} \\
- \frac{1}{V_4} \Tr \left\{ \frac{ i }{ i \dslash + i M_0 + i
\tilde{M}^{\alpha } \tau ^{\alpha } } \tau ^{\beta } \right\} &-&
I_F^\beta (M_0,M) \; + \;
\left( G^{-1}\right )^{\beta \gamma } \tilde{M} ^{\gamma } \; = \; 0 \; ,
\label{eq:42}
\ena
where
\bea
I_F &=& \frac{1}{V_4} \Tr \left\{ \frac{ i }{ i \dslash
+ i \mu \gamma_0 + i M_0 + i
\tilde{M}^{\alpha } \tau ^{\alpha } } \right\}
- \frac{1}{V_4} \Tr \left\{ \frac{ i }{ i \dslash +
i M_0 + i \tilde{M}^{\alpha } \tau ^{\alpha } } \right\}
\label{eq:43} \\
I_F^\beta &=& \frac{1}{V_4} \Tr \left\{ \frac{ i }{ i \dslash
+ i \mu \gamma_0 + i M_0 + i
\tilde{M}^{\alpha } \tau ^{\alpha } } \tau ^{\beta } \right\}
- \frac{1}{V_4} \Tr \left\{ \frac{ i }{ i \dslash +
i M_0 + i \tilde{M}^{\alpha } \tau ^{\alpha } } \tau ^{\beta } \right\} \; .
\label{eq:44}
\ena
One easily verifies that both functions $I_F$ and $I_F^\beta $ are
finite and need no regularization. In order to calculate these
functions, we first perform the color trace by introducing the
eigenvectors of the matrix $M^\alpha \tau ^\alpha $ defined in
(\ref{eq:12}). After taking the trace over Dirac indices,
the function $I_F$ becomes
\be
I_F(M_0,M) \; = \; \int \frac{ d^{4}k }{ (2\pi )^4 } \;
\left[ \frac{ 4 ( M_0 + iM ) }{ (k_0 + i \mu )^2 + (a+ib)^2 } \, - \,
\frac{ 4 ( M_0 + iM ) }{ k_0 ^2 + (a+ib)^2 } \right] \;
+ \; \left( M \rightarrow -M \right) \; ,
\label{eq:45}
\en
where the functions $a$ and $b$ depend on the momentum $\vec{k}$
as defined in (\ref{eq:24},\ref{eq:25}).
It is now straightforward to evaluate the $k_0$ integration
in (\ref{eq:45}):
\be
I_F(M_0,M) \; = \; - i \int _{ a( \vec{k}^2 ) < \mu }
\frac{ d^{3}k }{ (2\pi )^3 } \;
\frac{ 2 (M_0 + iM) }{ ia - b } \; + \;
\left( M \rightarrow -M \right) \; .
\label{eq:46}
\en
One observes that the color trace again enforces that
the function $I_F$ be real. Analogous considerations hold for the function
$I_F^\beta $. The final result for the gap equations
(\ref{eq:41},\ref{eq:42}) is
\bea
G_0^{-1} (M_0 - m) &=& M_0 \, I_+(M_0,M) \; - \; \frac{2}{\pi ^2 }
\int_0^{k_f} dk \; k^2 \; \frac{ M_0 a + M b }{ a^2+b^2 } \, ,
\label{eq:47} \\
G_c^{-1} M &=& M \, I_-(M_0,M) \; - \; \frac{2}{\pi ^2}
\int_0^{k_f} dk \; k^2 \; \frac{ Ma - M_0 b }{ a^2+b^2 } \, ,
\label{eq:48}
\ena
where the functions $I_{\pm }$ are defined in (\ref{eq:31a})
and $a(\vec{k}^2)$ and $b(\vec{k}^2)$ are defined in
(\ref{eq:24}) and (\ref{eq:25}) respectively.
The upper bound of the integration over the three-momentum $\vec{k}$
is provided by the Fermi sphere of radius $k_f$ where $k_f$ is
defined by $a(k_f^2)= \mu $.
The solutions of the coupled system (\ref{eq:47},\ref{eq:48}) provide the
constituent quark mass $M_0$ and the mass $M$ in the colored channel
as a function of the chemical potential $\mu $.
For physical applications, it is convenient to express the chemical
potential in terms of the baryonic density defined by
\be
\rho _B \; := \; -i \left( \frac{ \partial \ln Z }{ \partial \mu } (\mu )
\, - \, \frac{ \partial \ln Z }{ \partial \mu } (\mu =0) \right) \; .
\label{eq:49}
\en
{}From (\ref{eq:1}) and the Lagrangian ${\cal L}_D$ at finite chemical
potential $\mu $ in (\ref{eq:40}), one obtains
\be
\frac{ \partial \ln Z }{ \partial \mu } (\mu ) \; = \;
\Tr \left\{ \frac{ -1 }{ (k_0+i\mu) \gamma_0 + \vec{k}
\vec{\gamma } + i (M_0 + iM) } \gamma _0 \right\} \; .
\label{eq:50}
\en
A calculation along the line sketched above yields the simple
result
\be
\rho _B \; = \; 2 \int _{ a( \vec{k}^2 ) < \mu }
\frac{ d^{3}k }{ (2\pi )^3 } \; .
\label{eq:51}
\en
This provides the familiar relation $k_f^3 = 3\pi ^2 \rho _B$ verifying
that this relation also holds
in the context of the augmented model (\ref{eq:1}).

We have studied the solutions $M_0$ and $M$ of the gap equations
(\ref{eq:47},\ref{eq:48}) as a function of the Fermi momentum $k_f$.
The result is shown in Figure \ref{fig:6}.
The coupling strengths $G_0$ and $G_c$ were chosen in order for the system
to be in the confined phase $(M \not=0 )$ at zero density. For
increasing Fermi momentum $k_f$ one observes an increase in $M$
up to a critical momentum $k_f^c$, where $M$ rapidly drops to zero
implying that the deconfined phase is realized at high density.
The behavior of the color-singlet constituent quark mass $M_0$ is
different. It smoothly decreases and vanishes at $k_f^c$.
This behavior is different from that found at finite
temperature (i.e. section \ref{sec:5}) where the phase transition
causes a discontinuity of $M_0$ at the critical temperature.
The deconfining phase transition at finite density is accompanied
by the restoration of chiral symmetry as in the temperature case.

Figure \ref{fig:7} shows the critical color-triplet coupling
$G_c^{(crit)}$ as a function of the Fermi momentum $k_f$.
The dependence of $G_c^{(crit)}$ on the Fermi momentum $k_f$ is
qualitatively the same as the dependence on the temperature
(compare Figure \ref{fig:5}). The critical density is
nearly independent of the current mass of the quarks.

\section{Conclusions}
\bs

This paper describes how the NJL model can be modified minimally so as to
take into account
quark confinement. Confinement is understood in the sense that quark-anti-quark
thresholds which plague the standard NJL model
are screened by a random color background field:  physical
quantities are free of unwanted colored excitations. The mechanism that
confines the quarks
is analogous to Anderson localization in electronic systems and deconfinement
can be induced by temperature and/or density in a way that seems to be
consistent with QCD. In this model, deconfinement and chiral symmetry
restoration occur at the same critical point.
We have used, for simplicity, color $SU(2)$ group
but we have no reason to expect that qualitative
features would be modified if we were to consider the realistic $SU(3)$
gauge group. For applications to phenomenology, $SU(3)$ gauge group
will have to be treated. The extension to color $SU(3)$
remains to be worked out.

The problem that we hope to be able to resolve with the confining NJL
model is to describe how hadrons -- both mesons and baryons --
behave in medium at finite temperature and density. Since confinement is
suitably implemented in the model, we can treat the excitations that are not
amenable
to the conventional NJL model, such as scalar and vector mesons. For instance,
we should be able
to ``derive" the BR scaling predicted in mean field of effective
chiral Lagrangians \cite{br91} and address the properties of hadrons in hot
and dense medium created in heavy-ion collisions or compact
star matter \cite{newbr}. A microscopic model of the kind presented here
will have certain advantage over the ``macroscopic" treatment of \cite{br91}.
\newpage
\subsubsection*{Acknowledgments}
\bs

One of us (K.L) is indebted to J.\ Gasser for helpful
discussions on the analytic structure of Green's functions.
Part of this work was done while the authors were participating in
the 1995 Spring program on ``Chiral dynamics in hadrons and nuclei"
at the Institute for Nuclear Theory, University
of Washington, Seattle. We would like to thank the participants for
helpful discussions. We would also like to thank the INT for the hospitality
and the Department of Energy for partial support.

\bigskip
\section*{Appendices}
\renewcommand{\thesubsection}{\Alph{subsection}}
\renewcommand{\theequation}{\Alph{subsection}.\arabic{equation}}
\setcounter{subsection}{0}
\subsection{Some ingredients for the scalar correlation function }
\label{app:a}
\setcounter{equation}{0}
\bs

Here we calculate the functions $\Pi_s^0 (p^2)$, $K^{\alpha }(p^2)$
and $\Pi _s^{\alpha \beta }(p^2)$, defined in
(\ref{eq:e2},\ref{eq:e3},\ref{eq:e4}) that enter in the
scalar correlation function (\ref{eq:9}).
For this purpose, we need to evaluate the trace,
which extends over Lorentz- and color-space, of the functions
\be
\tr \left\{ S(k+p) S(k) \right\} \; , \hbo
\tr \left\{ \tau ^{\alpha } S(k+p) \tau ^{\beta } S(k) \right\} \; ,
\hbo
\tr \left\{ \tau ^{\alpha } S(k+p) S(k) \right\} \; ,
\label{eq:a1}
\en
where the fermion propagator $S(k)$, (\ref{eq:e5}),
\be
S(k) \; := \; \frac{ 1 }{ \kslash \, + \, i \left( M_0 +
i M^{\alpha } \tau ^{\alpha } \right) } \;
\label{eq:a2}
\en
possesses a non-trivial color structure. The Lorentz trace is
straightforward. The color trace is most conveniently evaluated by
introducing the eigenvectors $\vert \pm \rangle $ of the
matrix $M^\alpha \tau ^\alpha $ which form a complete set in
color space giving rise to a specific representation of the unit
element, i.e.
\be
1 \; = \; \sum _{i=\pm } \vert i \rangle \langle i \vert \; .
\label{eq:a3}
\en
For instance, the color trace of the first expression in (\ref{eq:a1})
is
\be
\sum _{i,l=\pm } \tr_L \left\{ \langle i \vert S(k+p)
\vert l \rangle \langle l \vert S(k) \vert i \rangle \right\} \; ,
\label{eq:a4}
\en
where $\tr _L $ indicates the trace over Lorentz indices only. Using
\be
S(k) \vert \pm \rangle \; = \; \frac{ 1 }{ \kslash \, + \, i \left( M_0
\pm i M^{\alpha } \right) } \, \vert \pm \rangle \; ,
\label{eq:a5}
\en
one obtains for (\ref{eq:a4})
\be
\tr _L \left\{ s(k+p) s(k) \right\} \; + \; (M \rightarrow -M) \; ,
\label{eq:a6}
\en
where
\be
s(k) \; = \; \frac{1}{ \kslash + i A } \;
\hbox to 2cm{\hfill with \hfill }
A = M_0 + i M \; .
\label{eq:a7}
\en
It is now straightforward to calculate $\Pi _s^0(p^2)$ in
(\ref{eq:e2}):
\be
\Pi _s^0 (p^2) \; = \; \frac{1}{G_0} \; - \; \int \frac{ d^4k }{ (2\pi )^4 }
\left\{ \frac{ \tr _L \, (\kslash + \pslash - iA)(\kslash -iA) }{
\left[ (k+p)^2 + A^2 \right] \, (k^2 + A^2) }
\; + \; (M \rightarrow -M) \right\} \; .
\label{eq:a8}
\en
The Lorentz trace can be easily performed. Introducing Feynman's
parametrization, one has
\be
\Pi _s^0 (p^2) \; = \; \frac{1}{G_0} \; - \; 4 \int _0^1 d\alpha \;
\int \frac{ d^4k }{ (2\pi )^4 }
\left\{ \frac{ k^2 + kp - A^2 }{ \left[ (k+ \alpha p)^2 + Q^2
\right]^2 } \; + \; (M \rightarrow -M) \right\} \; ,
\label{eq:a9}
\en
where $Q=\alpha (1- \alpha ) p^2 +A^2 $. After the substitution
$q= k + \alpha p$, we obtain the final result
\bea
\Pi _s^0 (p^2) &=& \frac{1}{G_0} \; - \; 4 \int _0^1 d\alpha \;
\int \frac{ d^4q }{ (2\pi )^4 }
\left\{ \frac{ q^2 - Q }{ \left[ q^2 + Q^2
\right]^2 } \; + \; (M \rightarrow -M) \right\} \; ,
\label{eq:a10} \\
&=& \frac{1}{ G_0 } \; - \; H_0(p^2) \; ,
\ena
with $H_0(p^2)$ defined in (\ref{eq:e8}).
Since the integral in (\ref{eq:a10}) is divergent, we cut off
the $q$-integral by a sharp O(4) cutoff $\Lambda $. This regularization
procedure is included into the definition of our low-energy effective
quark theory. We have ensured (see section \ref{sec:ch})
that chiral symmetry is not violated by the cutoff procedure.

For the evaluation of $K^\alpha (p^2)$, we proceed along the same line
as sketched above. We first perform the color trace in the last
expression of (\ref{eq:a1}):
\be
\sum _{i,l=\pm } \tr_L \left\{ \langle i \vert \tau ^\alpha S(k+p)
\vert l \rangle \langle l \vert S(k) \vert i \rangle \right\} \; = \;
\sum _{i,l=\pm } \tr_L \left\{ \langle i \vert \tau ^\alpha
\vert l \rangle s(k+p) s(k) \langle l \vert i \rangle \right\} \; .
\label{eq:a11}
\en
{}From (\ref{eq:12a}) and the orthogonality of the eigenvectors
$\vert \pm \rangle $, we find
\be
K^{\alpha }(p^2) \; = \; -i \frac{ M^\alpha }{M} \;
\int \frac{ d^4k }{ (2\pi )^4 }
\tr _L \left\{ s(k+p) s(k) \right\} \; - \; (M \rightarrow -M) \; .
\label{eq:a12}
\en
For the further calculation of $K^\alpha $, one might employ directly
the above results to finally obtain
\be
K^\alpha (p^2) \; = \;  - 4i \frac{ M^\alpha }{M} \int _0^1 d\alpha \;
\int \frac{ d^4q }{ (2\pi )^4 }
\left\{ \frac{ q^2 - Q }{ \left[ q^2 + Q^2
\right]^2 } \; - \; (M \rightarrow -M) \right\} \; ,
\label{eq:a13} \\
\en
and therefore $K^{\alpha } (p^2) = \frac{ M^\alpha }{M} H_v(p^2)$ with
$H_v(p^)$ defined in (\ref{eq:e9}).

It remains to show that $M^\alpha $ is an eigenvector of the
polarization matrix $\Pi _s^{\alpha \beta }$. To do this,
we first evaluate the color trace of
\be
\tr \left\{ \tau ^{\alpha } S(k+p) \tau ^{\beta } \frac{ M^\beta }{M}
S(k) \right\} \; = \;
\sum _{i,l=\pm } \tr_L \left\{ \langle i \vert \tau ^\alpha
\vert l \rangle s(k+p) s(k) \langle l \vert
\tau ^{\beta } \frac{ M^\beta }{M} \vert i \rangle \right\} \; .
\label{eq:a14}
\en
Since $\vert \pm \rangle $ are eigenvectors to $\tau ^\beta M^\beta $,
the expression (\ref{eq:a14}) is
\be
\frac{ M^\alpha }{M} \tr _L \left\{ s(k+p) s(k) \right\} \; + \;
(M \rightarrow -M) \; .
\label{eq:a15}
\en
This expression was already calculated to obtain $\Pi _s^0(p^2)$.
Since $M^\beta $ is also an eigenvector of the interaction matrix
$G^{\alpha \beta }$ (\ref{eq:18}), the final result is
\be
\Pi _s^{\alpha \beta } (p^2) \frac{ M^\beta }{M} \; = \;
\left( \frac{ 1}{ G_c} - H_0(p^2) \right) \;
\frac{M^\alpha }{M} \; .
\label{eq:a16}
\en


\subsection{ The equations of motion }
\label{app:b}
\setcounter{equation}{0}
\bs

In this section we derive an explicit expression for the
equations of motion (\ref{eq:7}) and (\ref{eq:8}).
To perform the color trace of the quark propagator $S(k)$, we
introduce the eigenstates $\vert \pm \rangle $ of the color
matrix $M^\alpha \tau ^\alpha $ (compare Appendix \ref{app:a}):
\be
\tr \, S(k) \; = \; \tr _L \, \sum _{i=\pm}
\langle i \vert \, S(k) \, \vert i \rangle \; ,
\label{eq:b1}
\en
where $\tr _L$ is the trace over Lorenz-indices only.
The equation of motion (\ref{eq:7}) becomes
\be
\frac{1}{G_0} (M_0-m) \; - \; i \int \frac{ d^4k }{ (2\pi )^4 } \;
\left\{ \frac{ \kslash - i A }{ k^2 + A^2 } \; + \;
(M \rightarrow -M) \right\} \; = \; 0 \; ,
\label{eq:b2}
\en
where $A$ was defined in (\ref{eq:a7}). It is straightforward
to evaluate (\ref{eq:b2}). One obtains
\be
\frac{1}{G_0} (M_0-m) \; - \; 8 M_0 \; \int \frac{ d^4k }{ (2\pi )^4 } \;
\frac{ k^2 + M_0^2 + M^2 }{ k^4 +2 (M_0^2 - M^2) k^2 +
(M_0^2+M^2)^2 }=0 \; .
\label{eq:b3}
\en
Setting $\tilde{M}^\alpha = i M^\alpha $,
we derive the explicit form of the remaining equation of motion
(\ref{eq:8}) in a similar fashion. The color-triplet part of the
quark propagator, \ $\tr \, S(k) \tau ^\alpha $, is
\be
\tr _L \sum _{i=\pm} \langle i \vert \, S(k) \tau ^\alpha \,
\vert i \rangle
\; = \; \tr _L \sum _{i=\pm} s(k) \langle i \vert \, \tau ^\alpha \,
\vert i \rangle \; = \; \tr _L s(k) \frac{ M^{\alpha } }{M}
\, - \, (M \rightarrow -M) \; ,
\label{eq:b4}
\en
with $s(k)$ from (\ref{eq:a7}). If $M^\alpha $ is an eigenstate
of the interaction matrix, i.e.\ $G^{\alpha \beta } M^{\beta }
= G_c \, M^\alpha $, the equation of motion (\ref{eq:8})
reduces to
\be
\frac{i}{G_c} M \; - \; i \, (-i) \int \frac{ d^4k }{ (2\pi )^4 } \;
\left\{ \frac{ \kslash - i A }{ k^2 + A^2 } \; - \;
(M \rightarrow -M) \right\} \; = \; 0 \; .
\label{eq:b5}
\en
A direct calculation yields
\be
\frac{1}{G_c} M \; - \; 8 M \; \int \frac{ d^4k }{ (2\pi )^4 } \;
\frac{ k^2 - M_0^2 - M^2 }{ k^4 +2 (M_0^2 - M^2) k^2 +
(M_0^2+M^2)^2 } \; .
\label{eq:b6}
\en
Both integrals in (\ref{eq:b3}) and (\ref{eq:b6}) can be expressed
with the help of the functions $I_0$ and $I_v$ defined, respectively, in
(\ref{eq:20}) and (\ref{eq:20a}), to obtain the desired result
presented in (\ref{eq:19}) and (\ref{eq:19a}).

\subsection{ Analytic structure of the scalar correlation function }
\label{app:b2}
\setcounter{equation}{0}
\bs

As argued in section \ref{sec:3.2}, we believe that the approximations
made to derive the scalar correlation function do not violate
any fundamental requirements of quantum field theory. Nevertheless
we observe a non-trivial analytic structure of the scalar
correlation function which might make one suspect that
some constraints may be incorrectly implemented. In this
subsection, we show by an explicit calculation that one particular
constraint, namely, the stability criterion, is indeed satisfied.
Of course there are many more constraints. Here we shall offer
one evidence that our model is compatible with a general axiom
of quantum field theory.
The investigation of further constraints which might provide
insight into the possible analytic structure of Green's functions
of a confining theory seems very interesting to us and will be relegated
to a future work.

For completeness, we rederive the stability criterion. Following
the standard procedure~\cite{ch84}, the correlation function of a local
operator $j(x)$ is written in momentum space as
\be
G(q) \; = \; \int d^4 x \; e^{iqx} \;
\langle 0 \vert \, T \, j(x) j(0)
\vert 0 \rangle \; ,
\label{eq:b2.1}
\en
where $T$ is the time-ordering operator. Rewriting the time ordering,
one has
\be
G(q) \; = \; \int d^4 x \; e^{iqx} \; \theta (x_0)
\langle 0 \vert \, [j(x), j(0) ] \,
\vert 0 \rangle \; + \; \int d^4 x \; e^{iqx} \;
\langle 0 \vert \,  j(0) j(x)
\vert 0 \rangle \;
\label{eq:b2.2}
\en
Inserting a complete set of eigenstates, the last expression becomes
\bea
\sum _n \int d^4 x && e^{i(q+p_n)x} \; \langle 0 \vert j(0)
\vert n \rangle \langle n \vert j(0) \vert 0 \rangle
\label{eq:b2.3} \\
&=& \sum _n (2 \pi )^4 \, \delta ^4 (q+p_n) \; \langle 0 \vert j(0)
\vert n \rangle \langle n \vert j(0) \vert 0 \rangle \; .
\nonumber
\ena
In the lab frame $(q^0 > 0)$, there is no contribution from the term
(\ref{eq:b2.3}) to the correlator (\ref{eq:b2.1}), if we are dealing
with a stable theory, since there are no states with negative energy
$(E_n=p_n^0 > 0)$. We therefore observe that the correlation
function vanishes for $x_0$ less than zero, i.e.
\be
\langle 0 \vert \, T \, j(x) j(0) \vert 0 \rangle \; = \;
f(x) \; \theta (x_0) \; .
\label{eq:b2.4}
\en
In the standard case of a constituent quark model, the stability
criterion is satisfied for the following reason: the correlation
function in momentum space is analytic in the upper
$q_0$ complex-plane and possesses cuts at the real axes due to
quark anti-quark thresholds (see Figure \ref{fig:b2.1} (a)).
For $x_0<0$ one might calculate the Fourier transform by closing
the path by a semi-circle in the upper half plane. Since there
are no poles or cuts, we conclude that the Fourier transformation
yields zero.

\bs
In order to check the stability criterion in our model, we first
study the analytic structure of the scalar correlation function
(\ref{eq:21a}). To this aim, we have explicitly calculated the
functions $H_{0/v}(p^2)$ in (\ref{eq:e8}) and (\ref{eq:e9}).
The result is
\bea
H_0 (p^2) &=& h(p^2;M_0,M) + h(p^2;M_0,-M) \; ,
\nonumber \\
H_v (p^2) &=& -i [ h(p^2;M_0,M) - h(p^2;M_0,-M) ]
\label{eq:b2.5} \\
\frac{ 4 \pi ^2 }{ p^2 } \, h(p^2; M_0,M) &=& \frac{\Lambda ^2}{p^2}
+ \frac{1}{2} \left( \frac{ 4A^2 }{ p^2 } +1 \right) ^{3/2}
\ln \frac{ \sqrt{ \frac{ 4A^2 }{ p^2 } +1 } + 1 }{
\sqrt{ \frac{ 4A^2 }{ p^2 } +1 } - 1 }
\nonumber \\
&-& \frac{1}{2} \left(
\frac{ 6 A^2 }{ p^2 } +1 \right) \ln \left( 1 +
\frac{ \Lambda ^2 }{ A^2 } \right)
\label{eq:b2.6} \\
&-& \frac{1}{2} \frac{ \left( \frac{ 4A^2 }{ p^2 } +1 \right) \left(
\frac{2 \Lambda ^2 + 4A^2 }{ p^2 } +1 \right) }{
\sqrt{ \left( \frac{ 4 \Lambda ^2 + 4A^2 }{ p^2 } +1 \right) } }
\ln \frac{ \sqrt{ \left( \frac{ 4 \Lambda ^2 + 4A^2 }{ p^2 } +1 \right) }
+1 }{ \sqrt{ \left( \frac{ 4 \Lambda ^2 + 4A^2 }{ p^2 } +1 \right) } -1 }
\; , \nonumber
\ena
where $A= M_0 + i M $. We confine the momentum to $\vert p \vert ^2
\le \Lambda ^2 $. In this case the only cuts which occurs in the
upper half $q_0$ complex-plane are shown in Figure \ref{fig:b2.1}(b).
Their orientation is given by the phase $\phi $ of the complex
number $M_0 + i M$. In order to perform the Fourier transform
to the coordinate space, we close the path in the upper half plane
by the contour depicted in Figure \ref{fig:b2.1}(b). It seems that
the Fourier transform does not yield zero as it would be
required by the stability criterion, since there are now the
contributions from the cut. Our main observation is, however, that
these contributions exactly cancel. In order to see this, we first
note that the correlation function $\Delta (q_0)$ possesses
reflection positivity, i.e.
\be
\Delta (z^{\ast }) \; = \; \Delta ^{\ast } (z) \; ,
\label{eq:b2.7}
\en
as the functions $H_{0/v}$ do. This can be shown either by a direct
inspection of (\ref{eq:b2.5}) or, which is more instructive, by
tracing it back to the fact that we consider color singlet correlation
functions. In the latter case, the appropriate superposition of
terms with $M$ and $-M$ provides (\ref{eq:b2.7}).
Equation (\ref{eq:b2.7}) immediately implies that
the complex part of the integration over the contour in the half-plane
vanishes, since this contour is chosen to be symmetric with
respect to the imaginary axis. In order to prove that the
contributions from the cuts cancel, it is sufficient to show that
the contribution from either cut is purely imaginary.
That this is indeed the case holds on very general grounds.
Although this may be well-known to the readers, we nevertheless feel that
we should give some arguments on this matter. We have to
prove that integrals of the type
\be
I := \int _{\cal C} dx \; K( \ln (x) ) f(x)
\label{eq:b2.8}
\en
produce purely imaginary results, where ${\cal C}$ is the contour
surrounding the cut (like the contour $ACB$ in Figure
\ref{fig:b2.1}(b)). In fact, one has to study the more general
integrand $ K[ \ln (g(x)) ] f(x)$, the integral of which, however,
can be traced back to the one in (\ref{eq:b2.8}) by a change
of variable $x \rightarrow y= g(x)$ and a redefinition of the
function $f(x)$. This change of variables provides us with a linear
cut in the complex plane. If we define the cut of the logarithm
to coincide with the negative real-axis, a straightforward calculation
yields
\be
I = \int _{-R}^{0} dx \; f(x) \; \left\{
K( \ln (-x) + i \pi ) \, - \, K( \ln (-x) - i \pi ) \right\} \; ,
\label{eq:b2.9}
\en
where $R$ is the length of the cut under considerations.
To continue we have to assume that the function $K(x)$ possesses a
Laurant-expansion around $x=0$. It is straightforward to check
that this is the case in the context of the scalar correlation
function of our model. We then have
\be
I = \sum _{n=-m}^{\infty } c_n \int _{-R}^{0} dx \; f(x) \; \left\{
( \ln (-x) + i \pi )^n \, - \, ( \ln (-x) - i \pi )^n \right\} \; .
\label{eq:b2.10}
\en
The even powers of $i \pi $ drop out and we end up with a purely
imaginary result. This completes the proof.

To summarize our results, the scalar correlation function
possesses a nontrivial analytic structure in momentum space,
e.g.\ consisting of cuts in the upper half $q_0$ complex-plane.
We have proven that the particular analytic structure is
compatible with the stability criterion (\ref{eq:b2.4}) due to
an intrinsic cancellation mechanism of the contributions from the cuts.

\bs
We finally present the scalar correlation function in Minkowski-space.
{}From the very beginning, our model is given in Euclidean space and
the Green's functions in Minkowski-space are defined by the Wick
rotation (see Figure \ref{fig:b2.2}). In our case, this Wick-rotation
is non-trivial, since we have to take into account the contribution
from the cuts in Figure \ref{fig:b2.2}. Suppose the scalar
correlation function in Euclidean space is part of some scattering
amplitude, e.g.
\be
S(p^2) \; = \; \int _{-\infty }^{\infty } dq_0^E \; \Delta (q^2) \,
{\cal G }(q^2,p^2)
\label{eq:b2.11}
\en
where ${\cal G}(q^2,p^2)$ is assumed to be analytic in the
first and third quadrants of the $q_0^E$-plane. From Figure
\ref{fig:b2.2} we have
\be
S(p^2) \; = \; \int _{-i\infty }^{i\infty } dq_0^E \; \Delta (q^2) \,
{\cal G }(q^2,p^2) \; + \; 2i \int _{-\infty }^{\infty } dx \;
I_{cut} \; ,
\label{eq:b2.12}
\en
where we have used that the contribution $I_{cut}$ from one cut is of
the form (\ref{eq:b2.10}) and purely imaginary ($I_{cut}$ is real).
Performing the substitution $q_0^M = -i q_0^E$ and redefining
$x=q_0^M$, we obtain
\be
S(p^2) \; = \; i \int _{-\infty }^{\infty } dq_0^M \; \Delta (-q^2) \,
{\cal G }(-q^2,p^2) \; + \; 2i \int _{-\infty }^{\infty } dq_0^M \;
I_{cut} \; ,
\label{eq:b2.13}
\en
and arrive at the scalar correlation function in Minkowski space
\be
\Delta ^M (q^2) \; = \; \Delta (-q^2) \; + \; 2 I_{cut}
\label{eq:b2.14}
\en
Note that the contribution from the cuts is real.
We expect $I_{cut}$ to contribute only a background to
$\Delta ^M (q^2) $, whereas the significant structure, such as
imaginary parts from thresholds and poles from particle states,
is produced by the Euclidean scalar correlation function at
negative momentum squared.

\subsection{ Reduction of the Bethe-Salpeter amplitude }
\label{app:c}
\setcounter{equation}{0}
\bs

In order to solve the Bethe-Salpeter equation (\ref{eq:h2}) for the
hidden color structure of the pion, one has to perform the color
trace of polarizations containing two quark propagators $S(k)$.
Since the calculational technique parallels very much that performed
to obtain the scalar correlation function, we only sketch the
derivation briefly and refer the reader to Appendix \ref{app:a}
for further details. The traces of interest are
\bea
\tr \{ \gamma _5 S(k+p) \gamma _5 S(k) \} &=&
\tr _L \{ \gamma _5 s(k+p) \gamma _5 s(k) \} \, + \,
(M \rightarrow -M) \; ,
\label{eq:c1}\nonumber \\
\tr \{ \tau ^\alpha \gamma _5 S(k+p) \gamma _5 S(k) \} &=&
\frac{M^{\alpha }}{M} \, \tr _L \{ \gamma _5 s(k+p) \gamma _5 s(k) \}
\, - \, (M \rightarrow -M) \; ,
\label{eq:c2} \nonumber\\
\tr \{ \tau ^\alpha \gamma _5 S(k+p) \tau ^\beta \gamma _5
S(k) \} \frac{M^\beta }{M} &=&
\frac{M^{\alpha }}{M} \, \tr _L \{ \gamma _5 s(k+p) \gamma _5 s(k) \}
\, + \, (M \rightarrow -M) \; \nonumber\\
\label{eq:c3}
\ena
with $s(k)$ defined in (\ref{eq:a7}). The next step is to
calculate the trace over Lorentz indices $\tr _L$:
\be
\tr _L \left\{ \gamma _5 \frac{ \kslash + \pslash - iA }{ (k+p)^2 + A^2 }
\gamma _5 \frac{ \kslash - iA }{ k^2 + A^2 } \right\}
\; = \; -4 \frac{ k^2 + A^2 + kp }{ \left[ (k+p)^2 +A^2 \right] \,
(k^2 + A^2 ) } \; .
\label{eq:c4}
\en
To obtain the desired matrix elements entering into (\ref{eq:h2}),
an integration over the loop momentum $k$ is required. Introducing
Feynman's parametrization and shifting the momentum integration
$q= k + \alpha p$ yields
\be
\Tr \left\{ \gamma _5 s(k+p) \gamma _5 s(k) \right\} \; = \;
-4 \int _0^1 d\alpha \; \int (q) \; \frac{ q^2 - \alpha (1 - \alpha)
p^2 +A^2 }{ \left[ q^2 + \alpha (1-\alpha ) p^2 + A^2 \right] ^2 }
\; .
\label{eq:c5}
\en
Inserting (\ref{eq:c5}) into (\ref{eq:c1}-\ref{eq:c3}), we have all
the ingredients to derive the result (\ref{eq:h4}).

\subsection{ Calculation of the electromagnetic form factor }
\label{app:d}
\setcounter{equation}{0}
\bs

Here we evaluate the matrix element (\ref{eq:h18}), which is
directly related to the electromagnetic form factor.
For this, we first perform the color trace by the method
employed several times before (for details see Appendix \ref{app:a}):
\be
(P_0^2 - P_1^2) \, W_\mu ^0 \; - \; 2 P_0 P_1 \, W_\mu ^v \; .
\label{eq:d1}
\en
The functions
\bea
W_\mu ^0 &=& - \int \frac{ d^4k }{ (2\pi )^4 } \; \tr_L \left\{
\gamma _5 \, s(k - \frac{p}{2}) \, \gamma _\mu \,
s(k + \frac{p}{2}) \, \gamma _5 \, s(k - \frac{p}{2}) \right\}
\; + \; (M \rightarrow -M) \; ,\nonumber\\
\label{eq:d2} \\
W_\mu ^v &=& + i \int \frac{ d^4k }{ (2\pi )^4 } \; \tr_L \left\{
\gamma _5 \, s(k - \frac{p}{2}) \, \gamma _\mu \,
s(k + \frac{p}{2}) \, \gamma _5 \, s(k - \frac{p}{2}) \right\}
\; - \; (M \rightarrow -M)\nonumber\\
\label{eq:d3}
\ena
are introduced for abbreviation. The propagator $s(k)$ is given by
(\ref{eq:a7}). Performing the Lorentz trace of the terms under
investigation in (\ref{eq:d2},\ref{eq:d3}), i.e.
\be
\frac{ 4 ( k_\mu + \frac{ p_\mu }{2} ) }{ \left[
(k- \frac{p}{2})^2 +A^2 \right]  \, \left[ (k+ \frac{p}{2})^2 +A^2
\right] } \; ,
\en
one obtains
\bea
W_\mu ^0 &=& \frac{1}{2} p_\mu \; \int \frac{ d^4k }{ (2\pi )^4 } \;
\frac{ 1  }{ \left[
(k- \frac{p}{2})^2 +A^2 \right]  \, \left[ (k+ \frac{p}{2})^2 +A^2
\right] }  \; + \; (M \rightarrow -M)
\label{eq:d5} \\
W_\mu ^v &=& - \frac{i}{2} p_\mu \; \int \frac{ d^4k }{ (2\pi )^4 } \;
\frac{ 1  }{ \left[
(k- \frac{p}{2})^2 +A^2 \right]  \, \left[ (k+ \frac{p}{2})^2 +A^2
\right] }  \; - \; (M \rightarrow -M)
\;. \label{eq:d6}
\ena
It is of a particular interest to expand these functions for small $p^2$
because they are required to normalize the pion
Bethe-Salpeter amplitude and to calculate the pion charge radius.
A direct calculation yields
\be
W_\mu ^{0/v} \; = \; p_\mu \, \left( F_{0/v} \; - \;
p^2 \, R_{0/v} \; + \; O(p^4) \right) \; ,
\label{eq:d7}
\en
where $F_{0/v}$ are defined in (\ref{eq:h8},\ref{eq:h9}) and
\bea
R_0 &=& \frac{1}{2} \int \frac{ d^4k }{ (2\pi )^4 } \;  \frac{ A^2 -
k^2 }{ \left( k^2 + A^2 \right)^4 } \; + \; (M \rightarrow -M) \; ,
\label{eq:d8} \\
R_v &=& -\frac{i}{2}\int \frac{ d^4k }{ (2\pi )^4 } \;  \frac{ A^2
- k^2 }{ \left( k^2 + A^2 \right)^4 } \; - \; (M \rightarrow -M) \; .
\label{eq:d9}
\ena
Inserting (\ref{eq:d7}) in (\ref{eq:d1}) leads to the desired matrix
element.

In order to obtain the electromagnetic form factor for
non-vanishing momentum in Minkowski space, an analytic continuation
of the Euclidean momentum squared to negative values is needed.
This analytic continuation is difficult from a technical point
of view, since the momentum integration in (\ref{eq:d5}, \ref{eq:d6})
can be performed only numerically. Introducing
\be
K_\pm = (k \pm \frac{p}{2})^2 = k^2 - \frac{ P^2}{4} \pm
k \dot p \; , \hbo A_\pm = M_0 \pm iM \; ,
\label{eq:d11}
\en
a term of interest is
\be
\frac{ 1 }{ \left[ K_- + A_+^2 \right] \, \left[ K_+ + A_+^2 \right] }
\; + \; (A_+ \rightarrow A_-) \; .
\label{eq:d12}
\en
We first remove the complex parts that enter via $A_\pm $, namely,
\be
\frac{ K_+ K_- + B_- (K_- + K_+) + B_-^2- 4 M_0^2 M^2 }{
\left[ K_- ^2 + 2 K_- B_- + B_+^2 \right] \,
\left[ K_+ ^2 + 2 K_+ B_- + B_+^2 \right] } \; ,
\label{eq:13}
\en
where $B_{\pm } = M_0^2 \pm M^2$.
Introducing the angle $\alpha $ between the Euclidean four-vectors $k$
and $p$, the crucial observation is that in (\ref{eq:13}), only the
terms quadratic in  the external momentum $p$ appear:
\bea
K_+K_- &=& (k^2 - \frac{p^2}{4})^2 - k^2 p^2 \cos ^2 \alpha \; , \\
K_+^2 + K_-^2 &=& 2 \left[
(k^2 - \frac{p^2}{4})^2 + k^2 p^2 \cos ^2 \alpha \right]
\label{eq:d14} \\
K_+ + K_- &=& 2\left( k^2 - \frac{ p^2}{4} \right) \; .
\label{eq:d14a}
\ena
It is now a simple matter  to perform the analytic continuation
$p^2 \rightarrow - p^2$. We finally present the result for
the functions $W_\mu ^0$, $W_\mu ^v$ in (\ref{eq:d5},\ref{eq:d6})
\bea
W_\mu ^0 &=& \frac{1}{8 \pi ^3} p_\mu \; \int dV \;
\frac{ \alpha _- + 2 T_- \beta + T_-^2 - 4 M_0^2 M^2 }{
\alpha _-^2 + 4 T_- \alpha _- \beta + 2 T_+^2 \alpha _+
+ 4 T_-^2 \alpha _- + 4 T_- T_+^2 \beta + T_+^4 } \; ,\nonumber\
\label{eq:d15} \\
W_\mu ^- &=& - \frac{1}{8 \pi ^3} p_\mu \; \int dV \;
\frac{ 4 M_0 M (\beta + T_-) }{
\alpha _-^2 + 4 T_- \alpha _- \beta + 2 T_+^2 \alpha _+
+ 4 T_-^2 \alpha _- + 4 T_- T_+^2 \beta + T_+^4 } \; \nonumber\\
\label{eq:d16}
\ena
where $dV = d \alpha \; \sin ^2 \alpha \, d(k^2) \, k^2 , \,
\alpha \in [0,\pi]$ and
\be
\alpha _{\pm } = (k^2 \pm \frac{p^2}{4})^2 \pm
k^2 p^2 \cos ^2 \alpha \; , \hbo \beta = k^2 - \frac{p^2}{4} \; ,
\hbo T_\pm = M_0^2 \pm M^2 \; .
\label{eq:d17}
\en
The momentum integration $k^2$ and the angle integral in
(\ref{eq:d15},\ref{eq:d16}) are left to a numerical
calculation.

\newpage
\centerline{\large \bf Figure Captions }
\vspace{1cm}
\begin{figure}[h]
\caption{ The color-singlet and color-triplet constituent quark masses
$M_0$ and $M$ for $ \frac{ G_c \Lambda ^2 }{ 2 \pi ^2 }=2 $ and
$m/\Lambda =0.01$ for the confining phase (solid and short dashed line)
and for the deconfined phase with $M=0$ (long dashed line). }
\label{fig:1}
\end{figure}

\begin{figure}[h]
\caption{ The critical color-triplet coupling $G_c$ as function of the
color-singlet coupling $G_0$ for different values of the current mass
$m$ in units of the cutoff $\Lambda $. }
\label{fig:2}
\end{figure}

\begin{figure}[h]
\caption{ The scalar correlation function as function of the
   Euclidean momentum transfer for different values of the
   current quark mass $m$ for $\frac{ G_0 \Lambda ^2 }{ 2 \pi ^2 }=1$
   and $\frac{ G_c \Lambda ^2 }{ 2 \pi ^2 }=2$.}
\label{fig:3}
\end{figure}

\begin{figure}[h]
\caption{ Graphical representation of the electromagnetic form factor
   (a) and the pion decay constant (b). }
\label{fig:ch1}
\end{figure}

\begin{figure}[h]
\caption{ The electromagnetic form factor $F(p^2)$ as function
 of the negative Euclidean momentum $-p^2$. }
\label{fig:ch3}
\end{figure}

\begin{figure}[h]
\caption{ Mass $m_\pi $ and decay constant $f_\pi $ of the pion
   as function of the current mass $m$ for
   $\frac{ G_0 \Lambda ^2 }{ 2 \pi ^2 }=1$
   and $\frac{ G_c \Lambda ^2 }{ 2 \pi ^2 }=2$.
   Also shown are the constituent quark masses $M_0$ and $M$. }
\label{fig:ch2}
\end{figure}

\begin{figure}[h]
\caption{ The pion dispersion relation as function of the
   Euclidean momentum transfer for zero current quark mass
   $m=0$ and for $\frac{ G_0 \Lambda ^2 }{ 2 \pi ^2 }=1$
   and $\frac{ G_c \Lambda ^2 }{ 2 \pi ^2 }=2$.}
\label{fig:3b}
\end{figure}

\begin{figure}[h]
\caption{ The pseudoscalar correlation function as function of the
   Euclidean momentum transfer for a current quark mass
   $m=0.01 \Lambda$ and for $\frac{ G_0 \Lambda ^2 }{ 2 \pi ^2 }=1$
   and $\frac{ G_c \Lambda ^2 }{ 2 \pi ^2 }=2$.}
\label{fig:3a}
\end{figure}

\begin{figure}[h]
\caption{ The color-singlet and color-triplet constituent mass
  as function of the temperature $T$. $M_0$, $M$, and $T$ in units
  of the cutoff $\Lambda $. }
\label{fig:4}
\end{figure}

\begin{figure}[h]
\caption{ The color-triplet critical coupling strength $G_c$ as
 function of the temperature $T$ for different values of the
 current mass $m$ in units of the cutoff. }
\label{fig:5}
\end{figure}

\begin{figure}[h]
\caption{ The constituent quark masses $M_0$ and $M$ as a function of the
Fermi momentum $k_f$ for $\frac{ G_0 \Lambda ^2 }{2 \pi ^2 }=1$ and
$\frac{ G_c \Lambda ^2 }{2 \pi ^2 }= 2$.
$M_0$, $M$ and $k_f$ in units of the cutoff $\Lambda $. }
\label{fig:6}
\end{figure}

\begin{figure}[h]
\caption{ The color-triplet critical coupling strength $G_c$ as
 a function of the Fermi momentum $k_f$ for different values of the
 current mass $m$ in units of the cutoff. }
\label{fig:7}
\end{figure}

\begin{figure}[h]
\caption{ The analytic structure of the Euclidean scalar correlation
   function in the complex $q_0$ plane for the standard NJL model (a) and
   the confining model (b). Cuts are indicated by dashed lines. }
\label{fig:b2.1}
\end{figure}

\begin{figure}[h]
\caption{ The Wick rotation to relate the scalar correlation function
   in Euclidean space to that in Minkowski space. }
\label{fig:b2.2}
\end{figure}
\end{document}